\documentclass[final]{pasj00}

\begin{document}

\Received{30 May 2007}
\Accepted{28 August 2007}
\Published{}
\SetRunningHead{E.~D.~Miller et al.}{Suzaku Observations of the North Polar Spur}

\title{Suzaku Observations of the North Polar Spur:  Evidence for\\ Nitrogen
Enhancement}

\author{Eric D. \textsc{Miller},\altaffilmark{1}
Hiroshi \textsc{Tsunemi},\altaffilmark{2}
Mark W. \textsc{Bautz},\altaffilmark{1}
Dan \textsc{McCammon},\altaffilmark{3}
Ryuichi \textsc{Fujimoto},\altaffilmark{4}\\
John P. \textsc{Hughes},\altaffilmark{5}
Satoru \textsc{Katsuda},\altaffilmark{2}
Motohide \textsc{Kokubun},\altaffilmark{6}
Kazuhisa \textsc{Mitsuda},\altaffilmark{4}
F. Scott \textsc{Porter},\altaffilmark{7}\\
Yoh \textsc{Takei},\altaffilmark{4,8}
Yohko \textsc{Tsuboi},\altaffilmark{9}
Noriko Y. \textsc{Yamasaki}\altaffilmark{4}}
\altaffiltext{1}{Kavli Institute for Astrophysics and Space Research, Massachusetts Institute of Technology, Cambridge, MA 02139, USA}
\altaffiltext{2}{Department of Earth and Space Science, Osaka University, Toyonaka, Osaka 560-0043}
\altaffiltext{3}{Department of Physics, University of Wisconsin, Madison, WI 53706, USA}
\altaffiltext{4}{Institute of Space and Astronautical Science (ISAS), Japan Aerospace Exploration Agency (JAXA),\\ 3-1-1 Yoshinodai, Sagamihara, Kanagawa 229-8510}
\altaffiltext{5}{Department of Physics and Astronomy, Rutgers University, Piscataway, NJ 08854-8019, USA}
\altaffiltext{6}{Department of Physics, University of Tokyo, 7-3-1 Hongo, Bunkyo-ku, Tokyo 113-0033}
\altaffiltext{7}{NASA/Goddard Space Flight Center, Greenbelt, MD 20771, USA}
\altaffiltext{8}{Netherlands Institute for Space Research (SRON), 
Sorbonnelaan 2, 3584 CA Utrecht, the Netherlands}
\altaffiltext{9}{Department of Physics, Chuo University, 1-13-27 Kasuga, Bunkyo-ku, Tokyo 112-8551}
\email{milleric@mit.edu}


\maketitle

\begin{abstract}
We present observations of the North Polar Spur (NPS) using the X-ray
Imaging Spectrometer (XIS) aboard the {\it Suzaku\/} X-ray satellite.  The
NPS is a large region of enhanced soft X-ray and radio emission projected
above the plane of the Galaxy, likely produced by a series of supernovae
and stellar winds from the nearby Sco-Cen OB association.  The exceptional
sensitivity and spectral resolution of the XIS below 1 keV allow
unprecedented probing of low-energy spectral lines, including C~\textsc{vi}
(0.37 keV) and N~\textsc{vii} (0.50 keV), and we have detected
highly-ionized nitrogen toward the NPS for the first time.  For this single
pointing toward the brightest 3/4 keV emission ($l = 26.8\arcdeg$, $b =
+22.0\arcdeg$), the best-fit NPS emission model implies a hot 
($kT \approx 0.3$ keV), collisional ionization equilibrium (CIE) plasma
with depleted C, O, Ne, Mg, and Fe abundances of less than 0.5 solar, but
an enhanced N abundance, with N/O = $(4.0^{+0.4}_{-0.5})$ times solar.  The
temperature and total thermal energy of the gas suggest heating by one or
more supernovae, while the enhanced nitrogen abundance is best explained by
enrichment from stellar material that has been processed by the CNO cycle.
Due to the time required to develop AGB stars, we conclude that this N/O
enhancement cannot be caused by the Sco-Cen OB association, but may result
from a previous enrichment episode in the solar neighborhood.
\end{abstract}

\section{Introduction}

The North Polar Spur (NPS) is a region of enhanced soft X-ray and radio
emission projected above the plane of the Galaxy, associated with the
bright radio continuum source called Loop I.  Early studies suggested that
Loop I is an old, nearby supernova remnant (SNR), with a shell of
non-thermal, polarized radio continuum emission surrounded by a shell of
slowly-expanding neutral gas
\citep{Berkhuijsen1971,Berkhuijsenetal1971,Sofueetal1974,Heilesetal1980}.
Early rocket-based X-ray observations supported the SNR view, with
detection of emission that could be explained by diffuse shock-heated
plasma
\citep{Bunneretal1972,Cruddaceetal1976,Iwan1980,Schnopperetal1982,Rocchiaetal1984}.
Additional work by \citet{deGeus1992} showed that the combined stellar
winds and supernovae from the centrally-located Scorpius-Centaurus OB
association are sufficient to sweep out shells in the ISM, forming the
loop-like structures we see.

The {\it ROSAT\/} mission provided an unprecedented X-ray view of the NPS, which
shows up clearly in the 3/4 keV all-sky map \citep{Snowdenetal1997}.  Using
these data, \citet{Egger1995} and \citet{EggerAschenbach1995} also suggest
that Loop I was produced by the continuous effects of supernovae and
combined stellar winds from the Sco-Cen association.  They conclude that
the bright NPS emission arises from the most recent supernova shock wave
heating the outer shell of the superbubble, at a distance of about 100 pc
from the Sun.  Notably, the authors propose an interaction between the Loop
I superbubble and the Local Hot Bubble, citing detection of X-ray
shadowing from a dense ring of H~\textsc{i} at this interface.  

A different scenario has been argued in a series of papers by
\authorcite{Sofue1977}
(\yearcite{Sofue1977,Sofue1984,Sofue1994,Sofue2000,Sofue2003}) and recently
by \citet{Bland-HawthornCohen2003}.  Under this model, the NPS is the
remnant of a starburst or explosion near the Galactic center 15 Myr ago and
is at a distance of several kpc.  This scenario is based largely on
morphological arguments, however, and it is contradicted by other
observations.  For example, \citet{MathewsonFord1970} detect an interstellar
polarization feature at a distance of about 100 pc which clearly follows
much of the northern and eastern parts of Loop I, including the NPS, with
the expected polarization orientation.  In addition, the H~\textsc{i}
features seen nearby appear to be due to an interaction of Loop I with the
Local Bubble, as previously described \citep{EggerAschenbach1995}.  These
results strongly favor the local NPS model, although the Galactic
center model cannot yet be ruled out.

Our understanding of the origin of the NPS hinges on the plasma conditions,
including temperature and metal abundance, which can be constrained by
current X-ray instrumentation.  However, X-ray observations along this line
of sight are complicated by the contribution of various emission sources
comprising the soft X-ray background (SXRB).  The Local Hot Bubble (LHB;
\cite{Snowdenetal1990} and references therein), Galactic halo, and
additional diffuse Galactic material produce thermal emission below 2 keV,
consistent with 0.1--0.2 keV plasma, and dominated by low-energy emission
lines.  Contributions from unresolved Galactic and extragalactic point
sources produce continuum emission across a broad energy band.  These
emission components are affected by differing amounts of intervening
absorption.  

Recent {\it XMM-Newton\/}/EPIC-MOS observations of three NPS pointings at
low latitude offer some insight, modeling the various line-of-sight
emission components and identifying a $kT \approx 0.25$ keV plasma with
bright emission lines and metal abundances of about 0.5 solar
\citep{Willingaleetal2003}.  This high-temperature emission is associated
with the enhanced NPS feature seen in the {\it ROSAT\/} maps, and it
supports the view that the emission arises from re-heating of a superbubble
shell.

The relative chemical abundances and the plasma conditions within the NPS
hold the key to its formation.  With {\it Suzaku\/} we are able to probe
the low-energy emission lines of C~\textsc{vi} (0.37 kev), N~\textsc{vi}
(0.43 keV), and N~\textsc{vii} (0.50 kev) in this region for the first time.
Thanks to an unprecedented combination of effective area and spectral
resolution from a CCD-based instrument, the XIS \citep{Koyamaetal2007} has
produced new results on a range of soft X-ray sources
\citep{Miyataetal2007,Fujimotoetal2007,Smithetal2007,Hamaguchietal2007}.
Here we present {\it Suzaku\/}/XIS observations of the NPS in an effort to
further constrain the temperature, abundance, and structure of the emitting
material.

Throughout this paper, errors associated with values in the text or tables
are at 90\% confidence for a single parameter, unless otherwise noted.
Error bars shown in figures are 1-$\sigma$.

%
%
%

\section{Observations and Data Reduction}

\begin{figure}
\FigureFile(\linewidth,\linewidth){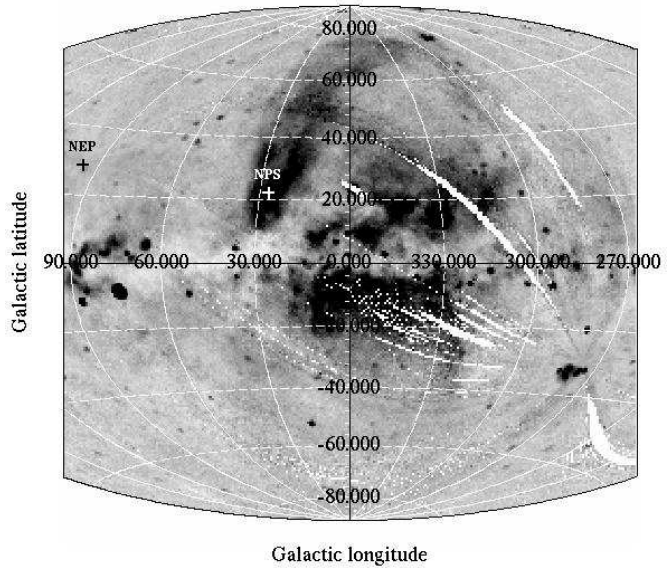}
\caption
{{\it ROSAT\/} 3/4 keV map toward the Galactic center, shown in inverse
grayscale \protect{\citep{Snowdenetal1997}}.  The location of our {\it
Suzaku\/} NPS pointing is indicated by the white cross.  The location of
the off-source NEP observation is shown by a black cross.}
\label{fig:rosat}
\end{figure}

The NPS was observed by the {\it Suzaku\/}/XIS during Science Working Group
(SWG) time on 3--4 October 2005 (ObsID 100038010), for a total integration
time of 46.1 ksec.  The pointing was chosen from the peak of the NPS
emission seen in the 3/4 keV {\it ROSAT\/} all-sky survey map
\citep{Snowdenetal1997}, toward $l = 26.8\arcdeg$, $b = +22.0\arcdeg$ (see
Figure \ref{fig:rosat}).  The observing parameters are listed in Table
\ref{tab:obs}.

\begin{table*}
\caption{Observing Parameters \label{tab:obs}}
\begin{center}
\begin{tabular}{lcc}
\hline
\hline
Target & North Polar Spur & North Ecliptic Pole$^{a}$ \\
Date   & 3--4 Oct 2005    & 2 Sep 2005 \\ 
(ObsID) & (100038010)      & (100018010) \\
       &                  & 12 Feb 2006 \\
       &                  & (500026010) \\
\hline
$\alpha$  & \timeform{17h22m21.0s} & \timeform{18h11m12.0s} \\ 
$\delta$  & \timeform{+04D45'24''} & \timeform{+66D00'00''} \\ 
$l$ & \phantom{+}\timeform{26D.84} & \phantom{+}\timeform{95D.77}  \\ 
$b$ & \timeform{+21D.96} & \timeform{+28D.67}  \\ 
$N_{\rm H}$ (cm$^{-2}$)$^{b}$ & $5.63\times 10^{20}$ & $4.3\times 10^{20}$ \\
\hline
\multicolumn{1}{l}{$t_{\rm exp,filt}$ (ksec)$^{c}$} \\
~~XIS0 & 35.7 & 136.7 \\
~~XIS1 & 39.9 & 142.3 \\
~~XIS2 & 38.4 & 136.4 \\
~~XIS3 & 38.7 & 137.9 \\
\hline
\multicolumn{3}{l}{\parbox{80mm}{\footnotesize
\footnotemark[$a$] This dataset was used as an off-source background.
\par\noindent
\footnotemark[$b$] \citet{DickeyLockman1990}
\par\noindent
\footnotemark[$c$] Effective exposure time of filtered datasets.
}}
\end{tabular}
\end{center}
\end{table*}

\subsection{Initial Data Preparation \label{sect:dataprep}}

The data were processed with rev0.7 of the XIS pipeline processing
software, producing a set of cleaned event lists with hot pixels removed
and including only ASCA grades\footnote{See 
{\it The\/} ASCA {\it Data Reduction Guide\/}
(http://heasarc.gsfc.nasa.gov/docs/asca/abc/abc.html).} 
0,2,3,4,6.  The recommended filtering criteria\footnote{See 
{\it The\/} Suzaku {\it Data Reduction Guide\/}
(http://heasarc.gsfc.nasa.gov/docs/suzaku/analysis/abc/).} 
were used to remove SAA transits and times of low geomagnetic cut-off
rigidity (COR), eliminating data with COR $<$ 6 GV.  In addition, all data
taken within $20\arcdeg$ of the sun-lit Earth limb or $10\arcdeg$ of the
dark Earth limb were excised to reduce contamination from scattered solar
X-ray flux.  Times of telemetry saturation were removed as well.

Analysis of the light curves revealed some times of high count rate for two
of the detectors, XIS0 and XIS1.  The high count rate in XIS0 lasted a
single orbit and was due to a high CCD temperature, the result of a
temporary thermal control system anomaly.  Housekeeping and count rate data
from the previous and subsequent orbits were normal, so only data during
the affected orbit were excised, eliminating 4.2 ksec of exposure time for
XIS0.
The high count rate in the back-illuminated (BI) chip XIS1 arises from
variations in the particle background as a result of varying geomagnetic
shielding (e.g., \cite{Snowden1998,Tawaetal2007}).  Rejection of particle
events is less effective for the BI sensor than the front-illuminated (FI)
sensors, and this produces a higher measured background.  Filtering based
on the cut-off rigidity can reduce the background, but at a cost of reduced
exposure time.  Further considerations of the particle background are
addressed in Section \ref{sect:particlebg}.  The combined filtering removed
22\% of XIS0 data, 13\% of XIS1 data, and 16\% of XIS2 and XIS3 data.  The
effective exposure times are shown in Table \ref{tab:obs}.

As described in Section \ref{sect:diffusespect}, we have used additional
{\it Suzaku\/} observations of the North Ecliptic Pole (NEP) as a
background field.  These observations were carried out on 2 September 2005
(ObsID 100018010) and 12 February 2006 (ObsID 500026010), also during {\it
Suzaku\/} SWG time, and have been presented by \citet{Fujimotoetal2007}.
The NEP data were processed in the same manner as the NPS data, using
rev0.7 software and identical filtering criteria.  The observations are
summarized in Table \ref{tab:obs}.

\begin{figure}
\FigureFile(\linewidth,\linewidth){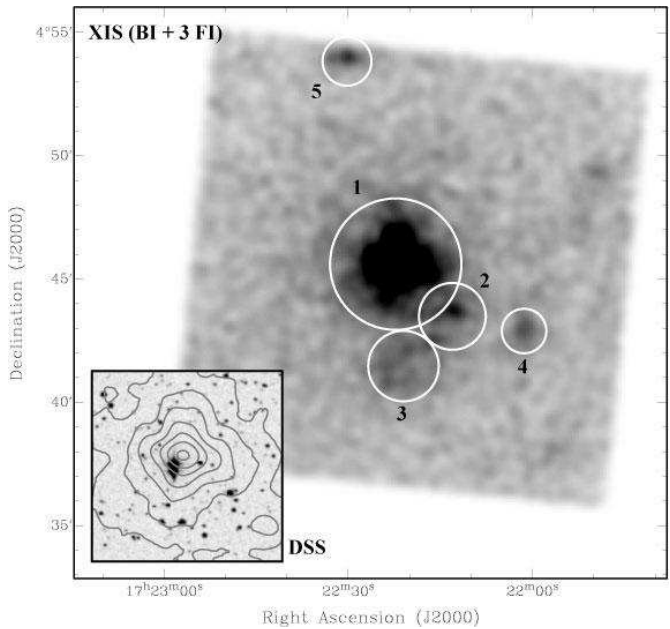}
\caption
{Combined 0.3-5.0 keV image of the XIS field of view, shown in inverse
grayscale.  Likely point sources are identified by white circles and are
discussed further in the text. The optical Digitized Sky Survey image of
the XIS field of view is shown as an inset, with the X-ray image
overplotted in contour.  The bright central X-ray point source (1RXS
J172223.9+044515) falls very close to a bright optical double star.}
\label{fig:image}
\end{figure}

Visual inspection of the binned broad-band image of the NPS field reveals a
number of likely point sources, including a bright source falling near the
center of the field (see Figure \ref{fig:image}).  To compromise between
removing all the point source flux and maximizing the diffuse emission
coverage, we used a smaller-than-normal masking/extraction region for the
bright central source (2.7\arcmin\ radius rather than the nominal
4.3\arcmin, 6 mm aperture for which the standard ARFs have been
constructed).  Four likely additional point sources were masked by hand.
See Section \ref{sect:point} for an analysis of these sources.

Spectra were extracted for each point source and for the portion of the NPS
field of view excluding these sources.  This latter ``point source
exclusion'' extraction was used as both the diffuse emission spectrum and
the background spectrum for the point sources.  The off-source background
spectrum from the NEP was extracted from an identical region projected in
detector coordinates.

\subsection{Contamination and Calibration Corrections}

Early in the {\it Suzaku\/} mission, it was discovered that the low-energy
effective area of each XIS instrument was deteriorating, likely due to a
build-up of contamination on the optical blocking filter of each sensor
\citep{Koyamaetal2007}.  Subsequent observations of calibration sources
have shown that the contamination is composed primarily of carbon with
small amounts of oxygen, and that the rate of deposition appears to be
decreasing.  In addition, the inferred thickness varies from sensor to
sensor, and the amount of contamination is non-uniform across each sensor,
with larger optical depth in the center \citep{Koyamaetal2007}.  The
effective area of the XIS at low energies is therefore dependent on chip
location, energy, and time.

To account for the contamination, we have used the empirical model
developed by the XIS calibration team.  This relies on regular
(approximately monthly) observations of point source calibration targets as
well as monthly accumulations of diffuse, bright Earth limb data.  We used
the {\it Suzaku\/} xissimarfgen tool (version 2006-10-26) to construct the
ancillary response functions (ARFs) for each spectral extraction region
discussed in this work.  This software corrects for the off-axis vignetting
as well as the spatial-, time- and energy-dependence of the contamination,
using the most reliable evolutionary model as of 2006-10-16, and assuming a
surface contaminant consisting solely of carbon and oxygen in the number
ratio $N_C/N_O = 6$ \citep{Ishisakietal2007}.  

We note that the contamination was discovered a month after the NPS
observations were completed, and the inferred contaminant thickness was
changing quite rapidly at that point in the mission (see Figure 14 of
\cite{Koyamaetal2007}; the NPS was observed 52 days after the XIS doors
opened on 2005 August 13).  The contamination was relatively small, with
central contaminant transmission at 0.5 keV (near the N~\textsc{vii} line)
ranging from 88\% (for the BI chip, XIS1) to 68\% (for XIS3, the worst FI
case).  Nevertheless, uncertainties in the contamination are a source of
systematic error in the broadband spectral modeling presented here.


The gain and spectral resolution of each CCD change on-orbit due to
radiation-induced decreases in the charge transfer efficiency.  To account
for this, we have constructed redistribution matrix functions (RMFs) using
the {\it Suzaku\/} xisrmfgen tool, version 2006-10-26.  These RMFs were
used in the spectral fitting discussed throughout this paper.

\subsection{Background Correction}

Spectral analysis of a diffuse source hinges on proper treatment of the
background, as parameters such as temperature can be affected by erroneous
modeling.  X-ray instruments in flight detect background counts from a
number of sources, and here we treat each one in some detail.

\subsubsection{Scattered Solar X-rays \label{sect:scatter}}

X-rays from the sun scatter off the outer parts of the Earth's
atmosphere, producing soft, field-filling emission which is
indistinguishable from a diffuse cosmic source (for a review see
\cite{SnowdenFreyberg1993}).  Two processes dominate: Thomson scattering
off of molecules and atoms, which mimics the line-rich solar spectrum at
low energies; and fluorescence scattering, which produces emission lines
from the primary atmospheric constituents, atomic and molecular nitrogen
and oxygen.  This contamination, especially from O K$\alpha$ fluorescence
near 0.53 keV, complicates spectral modeling near the energies we expect to
see carbon and nitrogen emission from the NPS and could produce false
detections of these lines.

\begin{figure}
\FigureFile(\linewidth,\linewidth){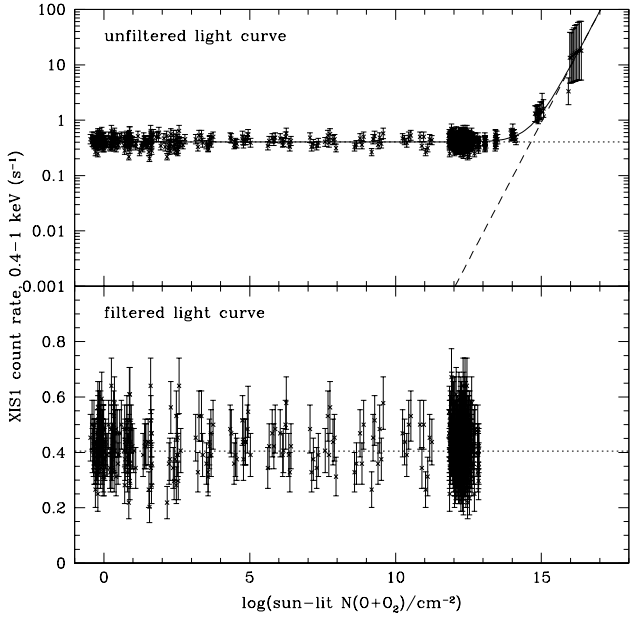}
\caption
{Count rate for XIS1 as a function of sun-lit oxygen column density, in the
0.4--1 keV range.  The oxygen column density includes contributions from
atomic and molecular states.  The top panel includes all X-ray events
without filtering for Earth angle, and scattered
solar flux clearly contaminates above $N({\rm O+O_2}) \approx 10^{14}$
cm$^{-2}$.  The solid line shows the best fit linear function described in
the text, with intercept due to the source flux (dotted line) and slope due
to the scattered solar flux (dashed line).  The bottom panel shows the
count rate on a linear scale, with data filtered as described in Section
\ref{sect:dataprep}.  All data used in this analysis were acquired with
sun-lit column densities below 10$^{13}$ cm$^{-2}$, and by extrapolating
the linear function, we expect scattered solar flux to contribute a few
percent of the count rate in this energy range.}
\label{fig:scatter}
\end{figure}

Since the intensity of scattered solar radiation depends on the column
density of the residual sun-lit atmosphere, the count rate is modulated
over the orbit.  Limits on solar contamination can be inferred from
measuring the soft X-ray count rate as a function of the sun-lit column
density along the line of sight.  Using the NRLMSISE-00 empirical model of
the atmosphere \citep{Piconeetal2002}, we have estimated the total sun-lit
O$_2$ plus O column along the telescope line of sight during the NPS
observation.  This is plotted in Figure \ref{fig:scatter} against the count
rate in the XIS1 0.4--1 keV band, in which we expect oxygen fluorescence to
produce the majority of flux.  The top panel shows the count rate with no
filtering based on Earth angle, and the rate clearly increases when the
total oxygen column exceeds $10^{14}$ cm$^{-2}$.  These high column density
observations are eliminated with the standard Earth angle filtering, as
shown in the bottom panel of Figure \ref{fig:scatter}.

To estimate the magnitude of solar X-ray contamination, we follow the
method used by \citet{Smithetal2007} in their {\it Suzaku\/} shadowing
observations toward the MBM 12 molecular cloud.  We fit a linear function
to the top panel of Figure \ref{fig:scatter} for $N({\rm O+O_2}) =
0$--$10^{17}$ cm$^{-2}$, obtaining a best fit count rate of
$(0.404\pm0.003) + (9.2\pm0.1)\times10^{-16} N({\rm O+O_2})$ cts
s$^{-1}$ (1-$\sigma$ errors).  Here the slope provides an estimate of
the scattered solar X-ray flux, while the intercept represents the count
rate from all other sources, including cosmic X-ray emission and the
particle background.  For comparison, the slope measured here is about 8
times larger than that for the MBM 12 observations \citep{Smithetal2007},
indicating a larger contamination from scattered solar X-ray flux at a
given sun-lit column density.

\begin{figure}
\FigureFile(\linewidth,\linewidth){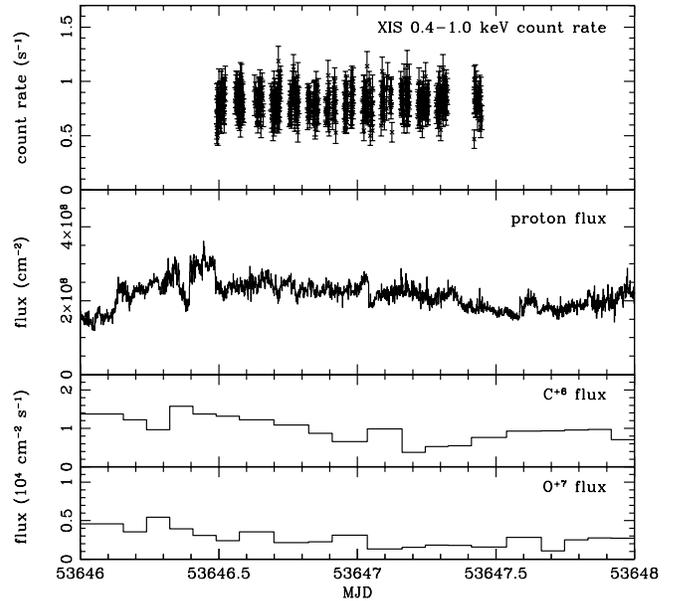}
\caption
{The top panel shows the combined 0.4--1.0 keV light curve for all four XIS
sensors.  The lower three panels show data collected by the {\it ACE\/}
satellite, corrected for travel time to the Earth, and used to estimate the
effects of geocoronal SWCX.  The second panel shows the proton flux, the
third panel shows the C$^{+6}$ ion flux, and the bottom panel shows the
O$^{+7}$ ion flux.  The time range was chosen to look for solar wind
enhancements of longer duration than our observations.  There is no
significant flare in the light curve, and no apparent solar wind flux
enhancements, so we conclude that geocoronal SWCX is unlikely to
contaminate our observations.}
\label{fig:ace}
\end{figure}

While the filtering on Earth angle eliminates column densities greater than
$5\times10^{12}$ cm$^{-2}$, about 50\% of the observing time occurs at
column densities between $5\times10^{11}$--$5\times10^{12}$ cm$^{-2}$,
where we expect solar contamination of a few percent.  We integrate the
linear slope over the good time intervals for the observation, estimating a
scattered solar count rate of 0.0012 cts s$^{-1}$ in the 0.4--1 keV
energy range.  This is equivalent to a flux of 0.02 cts s$^{-1}$
keV$^{-1}$ at 0.53 keV if all the emission is in the O K$\alpha$ line,
assuming an instrumental line width of 70 eV FWHM.  As discussed in Section
\ref{sect:diffuseother}, this amounts to a few percent of the count rate
in the N~\textsc{vii} and O~\textsc{vii} line region.  

\subsubsection{Solar Wind Charge Exchange}

Heavy ions in the solar wind undergo charge exchange (SWCX) with neutral
atoms in the Earth's geocorona and in the interplanetary medium.  The
resulting excited atoms release X-ray line emission which is generally
field-filling and indistinguishable from the more distant plasma emission
under study.  The solar wind heavy ion flux determines the X-ray line
strengths, and thus variations in the solar wind produce variations in the
contaminating flux on timescales of seconds (for geocoronal SWCX) to days
or longer (for interplanetary SWCX; e.g., \cite{Cravens2000}).

Mitigation strategies exist, as described by \citet{Fujimotoetal2007}, who
detect a short-term charge exchange enhancement in a long {\it Suzaku\/}
observation toward the NEP.  We follow the lead of these authors and
\citet{Smithetal2007}, and note the following.  First, the soft X-ray count
rate shows little variability with time, implying no charge exchange
variability on timescales less than 85 ksec (see Figure \ref{fig:ace}, top
panel).  Second, we have analyzed public data from the {\it Advanced
Composition Explorer\/} ({\it ACE\/}) satellite \citep{ACE}, which measures
solar wind properties, taken during our observation \citep{Rainesetal2005}.
The flux of protons, C$^{+6}$ ions, and O$^{+7}$ ions are shown in Figure
\ref{fig:ace}, corrected for the travel time from {\it ACE\/} to the Earth.
The proton flux varies by less than a factor of 1.4 during the {\it
Suzaku\/} observation, and it is about half the value seen during the NEP
flare, or nearly the quiescent value during that observation
\citep{Fujimotoetal2007}.  The ion fluxes are at least as low as the NEP
quiescent times.  We conclude that geocoronal SWCX is unlikely to
contaminate our data, but we cannot at this point constrain the effects of
SWCX in the interplanetary medium.

Since we use the NEP observations as an off-source background, we have
screened these data to remove the SWCX flare \citep{Fujimotoetal2007}.
There is no evidence for additional SWCX contamination in these
observations.  The final effective exposure time for the NEP data used as
background is shown in Table \ref{tab:obs}.

\subsubsection{Particle Background\label{sect:particlebg}}

All four sensors show count rate variations that are well-correlated with
the change in geomagnetic cut-off rigidity as the satellite orbits.  These
variations are similar to what is seen in observations dominated by the
particle background \citep{Tawaetal2007}.  A larger COR prevents energetic
particles from penetrating to the orbit of the satellite, reducing the
background count rate.  This is especially evident in the BI sensor, XIS1.
Due to its smaller depletion depth, the particle rejection efficiency of
XIS1 is lower than that of the FI chips, and it has a larger particle
background between 5--12 keV and a more noticeable count rate variation.
To reduce the background, we experimented with changing the minimum COR
value and analyzing the resultant light curves and spectra.  The default
constraint of COR $<$ 6 GV was found to be the best compromise between
minimum particle background and maximum exposure time, so no additional
filtering was performed.

Since the NPS and X-ray background emission sources are field-filling, we
must use non-contemporaneous observations to estimate the particle
background.  We used 800 ksec of data accumulated during observations of
the dark (night) Earth between 04 September 2005 and 21 May 2006
\citep{Tawaetal2007}.  These data were processed with rev0.7 of the
pipeline software, filtered for telemetry saturation and extracted using
the same detector regions as the science data to eliminate any spatial
dependence.  The background event rate is well-correlated with the COR, as
expected.  Since the NPS observations were taken at a different time, the
COR distribution and thus the integrated count rate will be different for
the accumulated night Earth data compared to the true particle background
during the science observations.  To correct for this, we have constructed
a COR-weighted background spectrum that comprises the same COR time
fractions as the NPS observations.  Similar COR weighting was performed on
the NEP data.

\subsubsection{Cosmic Background}

Galactic and extragalactic emission contributes to the background along the
line of sight.  Since it is difficult to separate these sources from the
extended NPS emission under study, we account for them in the spectral
modeling described in Section \ref{sect:diffusespect}.

\section{Analysis}

\subsection{Point-Like Sources \label{sect:point}}

The XIS field of view contains several point sources, including a bright
source located near the pointing center.  Two of these sources are found in
the {\it ROSAT\/} All-Sky Bright Source Catalog (\cite{Vogesetal1999};
sources 1 and 5 in Figure \ref{fig:image}), although they have no
classification.  The other sources are not associated with any known X-ray
source.

Source 1 is identified as 1RXS J172223.9+044515, a known X-ray source
within 30\arcsec\ of the XIS peak flux (and within the pointing error
circle of rev0.7 XIS data).  A search in other wavebands produces a known
double star near this location, HD 157310 ({\it B\/}=10.1,{\it V\/}=9.8)
and BD+04 3405B ({\it V\/}=10.8; see Figure \ref{fig:image}).  The fainter
star is within 6\arcsec\ of the cataloged position for the RXS source.

\begin{figure}
\FigureFile(\linewidth,\linewidth){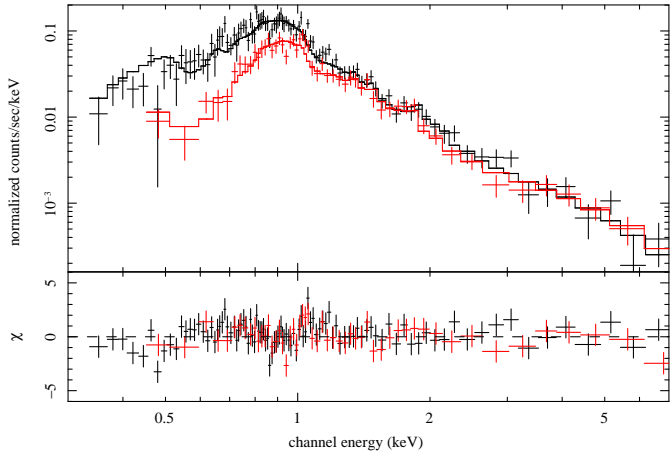}
\caption
{Spectra of the bright point source 1RXS J172223.9+044515 (Source 1 in the
text).  Black shows the BI spectrum and red shows the averaged FI spectrum.
Each is overplotted with the best-fit, APEC$+$power-law model.  The lower
panel shows the residuals.}
\label{fig:src1spec}
\end{figure}

The X-ray spectrum, shown in Figure \ref{fig:src1spec}, was fitted with a
number of models.  The background was provided by the full field of view,
excluding the point sources.  Absorbed single-emission-component models
(blackbody, thin thermal plasma, or power-law) are ruled out as poor fits.
Two-component models produce better results, with similar fits coming from
an absorbed thermal+power-law model and an absorbed thermal+thermal model
($\chi^{2}_r = 1.3$--1.5), using the APEC model \citep{Smithetal2001} for
the optically thin thermal emission component.  The model parameters are
listed in Table \ref{tab:src1params}; both models require a $kT \approx
0.75$ keV thermal plasma, assuming solar abundances.  The double thermal
model requires little line-of-sight absorption, while the thermal+power-law
model requires a solar-abundance absorbing column in excess of
$4.7\times10^{20}$ cm$^{-2}$, which is about 80\% of the expected Galactic
value along this line of sight.  The 0.5--5 keV flux of the source is 
$(5.8 \pm 0.2)\times 10^{-13}$ erg cm$^{-2}$ s$^{-1}$ (1-$\sigma$ errors).

\begin{table}
\begin{center}
\caption{Model Parameters for 1RXS J172223.9+044515\label{tab:src1params}}
\begin{tabular}{lrr}
\hline
\hline
                    & Model 1$^{\rm a}$ & Model 2$^{\rm b}$ \\
\hline
$kT_1$ (keV)       & 0.75 $\pm0.02$ & 0.75 $\pm0.02$ \\
$kT_2$ (keV)       & 3.1 $\pm0.3$  & ... \\
$\Gamma$            & ...       & 2.3 $\pm0.1$ \\
$N_{\rm H}$ ($10^{20}$ cm$^{-2}$) & $<$ 0.5       & $>$ 4.7 \\
$\chi_{\nu}^2$ (dof)      & 1.04 (488)    & 1.02 (488) \\
\hline
\multicolumn{3}{l}{\parbox{80mm}{\footnotesize
\footnotemark[$a$] Absorbed, two-APEC model.
\par\noindent
\footnotemark[$b$] Absorbed, APEC + power law model.
}}
\end{tabular}
\end{center}
\end{table}

It is likely this X-ray source is associated with one of the two bright
stars along this line of sight.  The brighter star was originally
classified as a G5 spectral type \citep{HDCat}, although later analysis
classified it as A7II/III \citep{Houk}.  The fainter star has no published
MK type.  From simple spectral parallax, and ignoring extinction, the
bright star would be about 88 pc distant if a G5 dwarf and 555 pc distant
if an A7III.  The unabsorbed 0.5--5 keV luminosities would be 
$L_X($88 pc$) = 5.4$--$6.3\times 10^{29}$ erg s$^{-1}$ and 
$L_X($555 pc$) = 2.2$--$2.5\times 10^{31}$ erg s$^{-1}$, depending on the true
absorbing column.

\begin{figure}
\begin{minipage}[t]{80mm}
\FigureFile(\linewidth,\linewidth){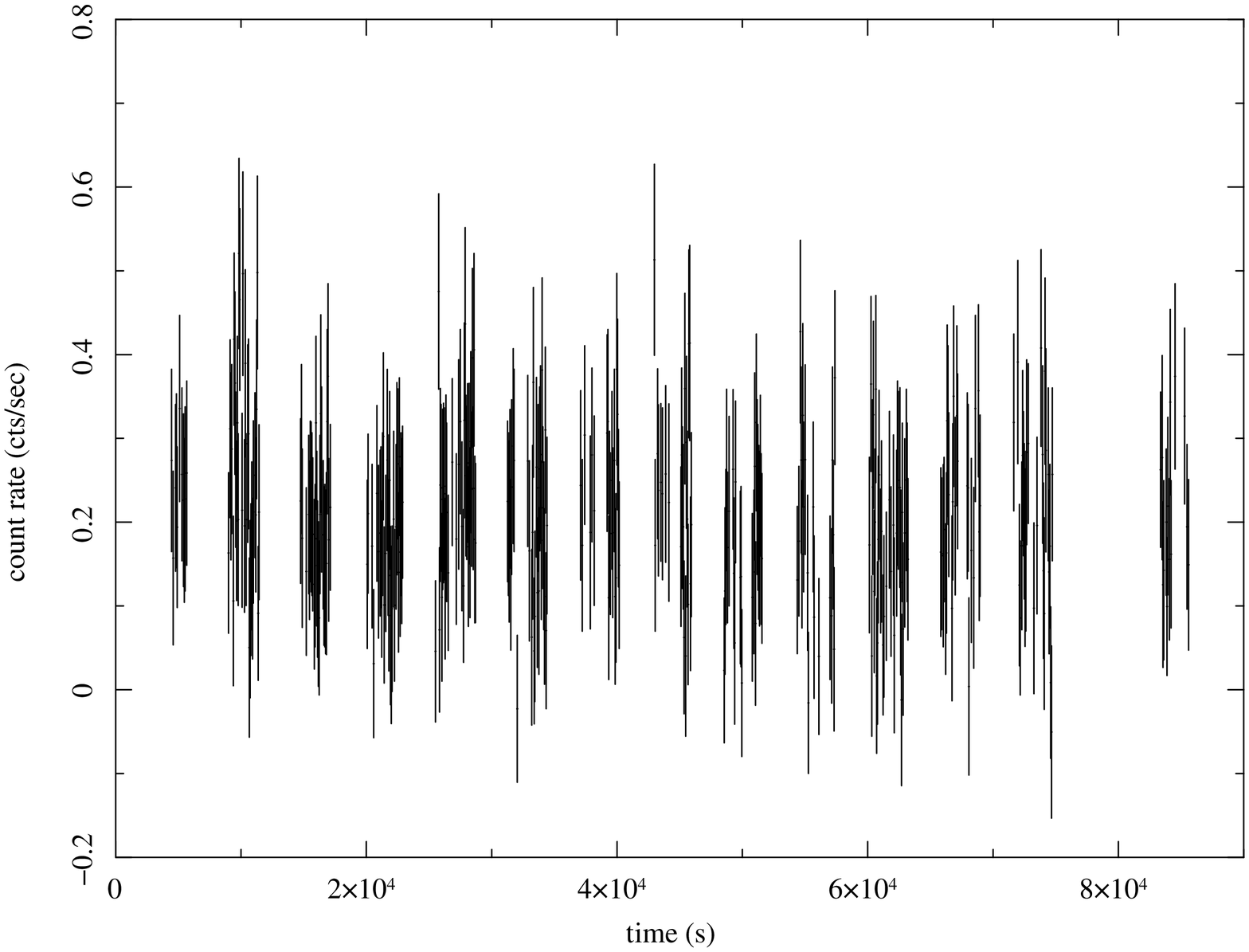}
\end{minipage}
\begin{minipage}[b]{80mm}
\FigureFile(\linewidth,\linewidth){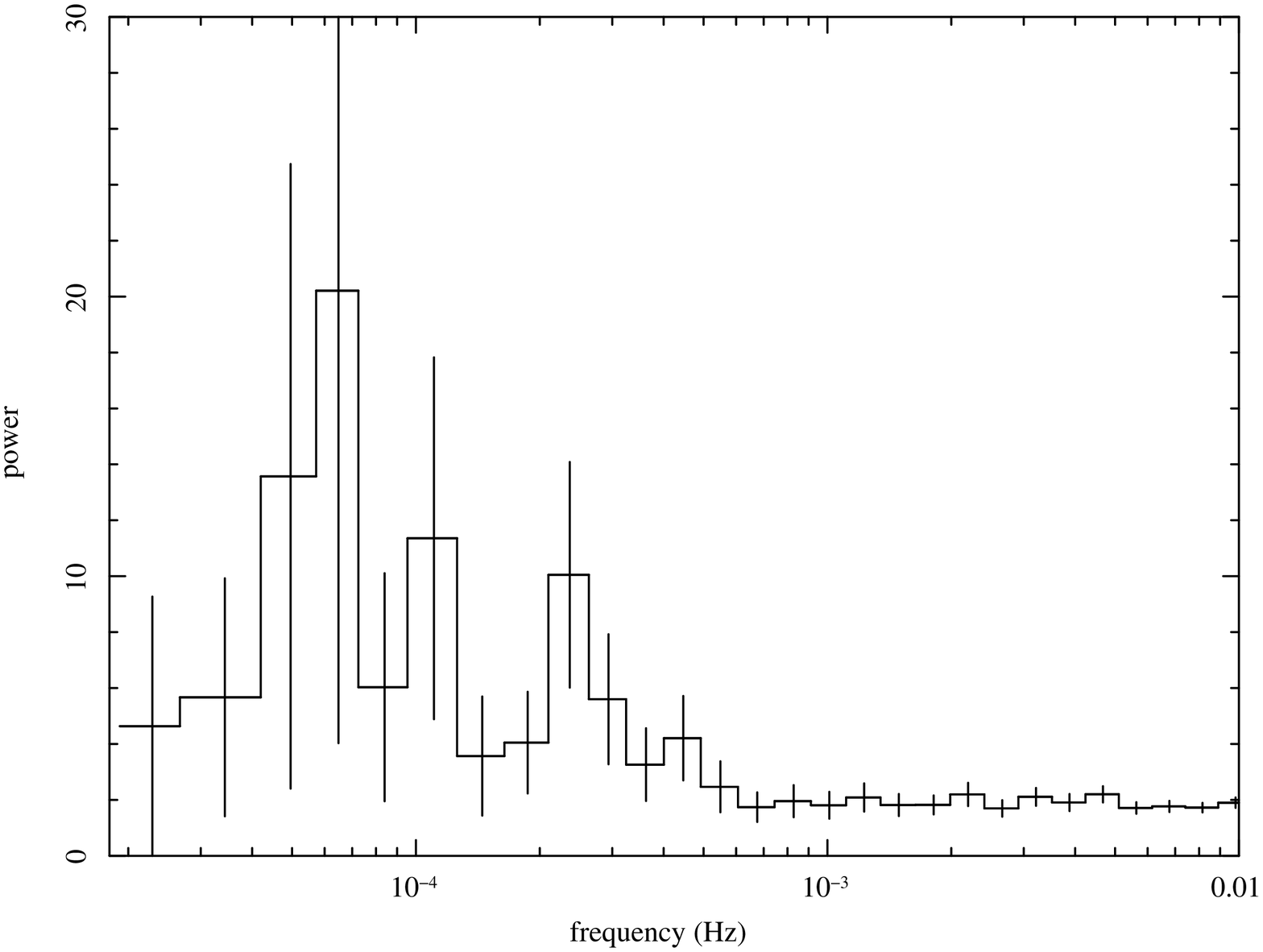}
\end{minipage}
\caption
{Background-subtracted light curve (top) and power spectrum (bottom) for
the bright central point source 1RXS J172223.9+044515.  Data from all four
sensors have been combined, spanning the 0.5--5.0 keV energy range.  A hint
of variability is seen, although the low signal-to-noise complicates
analysis.}
\label{fig:src1lc}
\end{figure}

To search for variability, we produced a background-subtracted light curve
and power spectrum from all four XIS sensors over the 0.5--5 keV band.
These are shown in Figure \ref{fig:src1lc}.  No variability is seen at
periods less than about 30 min, down to the time resolution of 8 sec.  A
peak appears around 75 min, and peaks with less significance are present at
longer periods.  There is also a hint of flaring in the light curve.  The
low signal-to-noise of the source and background prevent a more
quantitative timing analysis.

From the spectral model and likely optical counterpart, we tentatively
classify this source as an RS CVn interacting binary.  These systems
produce strong thermal coronal emission, they are common soft X-ray sources
in the Galaxy, and the derived X-ray luminosities span the range expected
for these objects (e.g., \cite{Drakeetal1989}).  The light curve is also
consistent with this conclusion, ruling out very large brightness
fluctuations but allowing flaring of less than 50\% or so over time scales
less than the 85 ksec observation window.  Note that the distance inferred
would place the source beyond the near edge of the assumed NPS ``bubble'',
and the large absorbing column implied by the thermal+power-law model is
consistent with results from fitting the diffuse soft X-ray emission (see
Section \ref{sect:diffusespect}).  Further classification of this source would
require detailed optical and UV spectroscopic analysis, which is beyond the
scope of this paper.

Source 2 is projected 3\arcmin\ from the central source, and has no known
X-ray counterpart within the expected pointing error circle.  It lies
30\arcsec\ from a known radio source, RC J1722+0442, which has been
identified from broad-band optical spectroscopy as a $z \approx 1$ radio
galaxy \citep{Verkhodanovetal2002}.  The extracted spectra for this source
are of low S/N, and therefore we have not performed additional analysis.
Sources 3 and 4 are very faint, although they appear to be discrete
emission.  There are no counterparts for these objects in the literature.
Source 5 is within 30\arcsec\ of the {\it ROSAT\/} object 1RXS
J172228.9+045402, although it falls at the very edge of the field of view,
making analysis difficult.  

\subsection{Diffuse Emission \label{sect:diffuse}}

\subsubsection{Spectral Modeling \label{sect:diffusespect}}

The line of sight to the NPS probes a number of foreground
and background sources which emit soft X-ray radiation.  The degree
of absorption expected in each component depends on the particular
schematic model invoked, of which there are several in the literature
\citep{Iwan1980,Davelaaretal1980,deGeus1992,EggerAschenbach1995,Willingaleetal2003}.
The models invariably include an unabsorbed Local Hot Bubble plasma that
extends 50--100 pc from the Sun, depending on the pointing direction.  This
is terminated by a dense H~\textsc{i} sheet inferred from differential
absorption in spectra of stars at known distance
\citep{Iwan1980,CenturionVladilo1991,EggerAschenbach1995}.  This sheet
encloses the Loop I (and NPS) bubble and has a column density of
$10^{20}$--$10^{21}$ cm$^{-2}$.  The NPS bubble fills some portion of the
Loop I bubble, beyond which lies the residual Galactic H~\textsc{i} column,
the Galactic halo plasma, and any extragalactic diffuse and unresolved
point-source emission.

To allow direct comparison to the {\it XMM-Newton\/} data, we have adopted
the spectral model components of \citet{Willingaleetal2003}, who observed
three fields towards the NPS at slightly different Galactic latitudes with
the EPIC-MOS.  This model consists of an unabsorbed, solar-abundance Local
Hot Bubble (LHB) with fixed $kT = 0.1$ keV; a single-temperature,
solar-abundance Galactic halo (GH) component with fixed $kT = 0.1$ keV,
subject to the full Galactic absorbing column; a broken power-law
extragalactic background (EB) to account for unresolved point sources, with
photon indices $\Gamma(<0.7~{\rm keV}) = 2.0$ and 
$\Gamma(>0.7~{\rm keV}) = 1.4$, subject to the full Galactic absorbing
column; and a thermal NPS component with freely-varying $kT$, abundances,
and $N_{\rm H}$.  The LHB, GH, and EB components together form the X-ray
background (XRB), and we allow only the normalizations of these components
to vary.  The shape of the EB power-law is identical to that used by
\citet{Willingaleetal2003}, and it includes a steepening at softer energies
to account for unresolved Type I AGN (e.g.,
\cite{RobertsWarwick2001,Baueretal2004,DeLucaMolendi2004}).  In the
remainder of this paper, we refer to this set of spectral components as the
``local background model'', since it accounts for the background directly
along the NPS sightline.

The model fitting was performed on all four XIS spectra simultaneously
using XSPEC v.12.3.1 \citep{Arnaud1996}.  The APEC model
\citep{Smithetal2001} was used for thin thermal plasma components (LHB, GH,
and NPS), with freely-varying C, N, O, Ne, Mg, and Fe abundances for the
NPS.  We fixed the Galactic $N_{\rm H}$ absorbing column to $5.63\times
10^{20}$ cm$^{-2}$ \citep{DickeyLockman1990}, using this as a maximum
allowable value for the NPS component.  We perform the fitting over the
0.4--3 keV energy range for the FI detectors and 0.3--3 keV for the BI
detector.  While the XIS BI device has measurable sensitivity between
0.2--0.3 keV (the C-band), due to uncertainties in the XIS calibration and
deficiencies in the APEC model (R.~Smith, private communication) in this
energy range, we have excluded it from our analysis.  An accumulated,
COR-weighted spectrum of the dark Earth was used to subtract the particle
background, as described in Section \ref{sect:particlebg}.  

As a check, we performed a model fit with an off-source pointing to
characterize the background, using the two long SWG observations of the NEP
\citep{Fujimotoetal2007}.  While this pointing is $62\arcdeg$ away from the
NPS and at a very different Galactic longitude ($96\arcdeg$ for the NEP
vs.~$27\arcdeg$ for the NPS), it is at a similar Galactic latitude
($29\arcdeg$ vs.~$22\arcdeg$; see Figure \ref{fig:rosat}).  As such, it
provides a first-order removal of the X-ray background components (LHB, GH,
EB) and the particle background (after correcting to a similar COR
distribution; see Section \ref{sect:particlebg}).  We note that the
Galactic absorbing column is
somewhat lower toward the NEP ($4.3\times 10^{20}$ cm$^{-2}$) than the NPS
($5.6\times 10^{20}$ cm$^{-2}$; \cite{DickeyLockman1990}), and we do not
take this into account.  The NEP spectrum is subtracted from the NPS
spectrum during fitting, and we fit only a single absorbed APEC component
representing the NPS plasma.  The absorbing $N_{\rm H}$ is allowed to vary
up to the maximum value of $5.6\times 10^{20}$ cm$^{-2}$, and the plasma
temperature, normalization, and C, N, O, Ne, Mg, and Fe abundances are also
allowed to vary.  In the remainder of the paper, we refer to this model as
the ``off-source background model''.

\begin{table}[t]
\begin{center}
\caption{Diffuse Emission Model Parameters \label{tab:models}}
\begin{tabular}{lcc}
\hline
\hline
background model      & local$^{a}$ & off-source$^{b}$ \\
\hline
\multicolumn{3}{l}{extragalactic backgr.} \\
\hline
$N_{\rm H}/N_{\rm H,tot}^{c}$ & 1 & ... \\
$\Gamma_1$ & 2.0 & ... \\
$\Gamma_2$ & 1.4 & ... \\
$E_{\rm break}$ (keV) & 0.7 & ... \\
norm at 1 keV & 7.5 $\pm$ 0.4 & ... \\
~~~(ph/cm$^{2}$/s/keV/sr) & & \\
\hline
\multicolumn{3}{l}{Local Hot Bubble} \\
\hline
$N_{\rm H}/N_{\rm H,tot}^{c}$ & 0 & ... \\
$kT$ (keV) & 0.1 & ... \\
abundance$^d$ & solar & ... \\
EM (pc cm$^{-6}$) & $<$ 0.0029 & ... \\
\hline
\multicolumn{3}{l}{Galactic halo} \\
\hline
$N_{\rm H}/N_{\rm H,tot}^{c}$ & 1 & ... \\
$kT$ (keV) & 0.1 & ... \\
abundance$^d$ & solar & ... \\
EM (pc cm$^{-6}$) & 0.056 $^{+0.006}_{-0.007}$ & ... \\
\hline
\multicolumn{3}{l}{North Polar Spur} \\
\hline
$N_{\rm H}/N_{\rm H,tot}^{c}$ & $>$0.71 & $>$0.97 \\
$kT$ (keV) & 0.29 $^{+0.01}_{-0.01}$ & 0.26 $^{+0.01}_{-0.01}$ \\
C$^{d}$  & $<$ 0.06                & $<$ 0.07 \\
N$^{d}$  & 1.33 $^{+0.52}_{-0.38}$ & 0.55 $^{+0.10}_{-0.09}$ \\
O$^{d}$  & 0.33 $^{+0.10}_{-0.07}$ & 0.16 $^{+0.01}_{-0.01}$ \\
Ne$^{d}$ & 0.51 $^{+0.16}_{-0.12}$ & 0.27 $^{+0.03}_{-0.03}$ \\
Mg$^{d}$ & 0.46 $^{+0.16}_{-0.12}$ & 0.33 $^{+0.07}_{-0.06}$ \\
Fe$^{d}$ & 0.50 $^{+0.13}_{-0.09}$ & 0.33 $^{+0.04}_{-0.03}$ \\
EM (pc cm$^{-6}$) & 0.089 $^{+0.031}_{-0.023}$ & 0.18 $^{+0.01}_{-0.01}$ \\
\hline
$\chi^2_r$ (dof) & 1.24 (961) & 1.21 (964) \\
\hline
\multicolumn{3}{l}{\parbox{85mm}{\footnotesize
\footnotemark[$a$] The ``local'' background model fits all emission
components along the NPS line of sight simultaneously.  Only a particle
background produced from night Earth observations is subtracted.
\par\noindent
\footnotemark[$b$] The ``off-source'' background model uses the NEP field
to subtract the EB, LHB, GH, and particle background components during
spectral fitting.
\par\noindent
\footnotemark[$c$] The ratio of the absorbing column to the full Galactic
column along this line of sight, $5.63\times{20}$ cm$^{-2}$
\citep{DickeyLockman1990}.
\par\noindent
\footnotemark[$d$] Abundances are given in terms of the solar value, as 
tabulated by \citet{AndersGrevesse1989}.
}}
\end{tabular}
\end{center}
\end{table}

\begin{figure}
\FigureFile(\linewidth,\linewidth){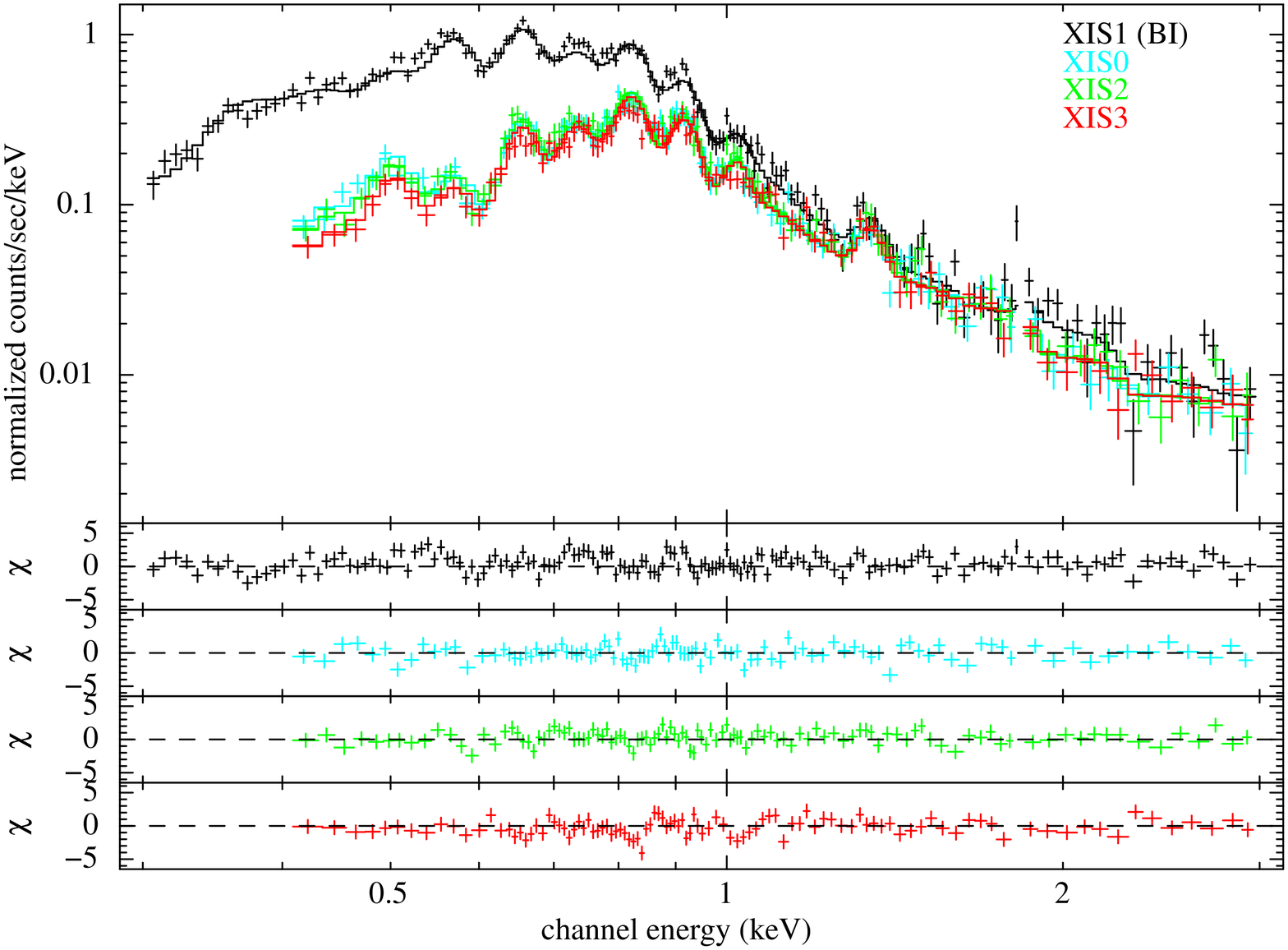}
\caption
{{\it Suzaku\/}/XIS spectra of the NPS overlaid with the best-fit local
background model.  The particle background has been subtracted from the
count rate spectra.  Black is used for the BI spectrum (XIS1), red, green
and cyan for the FI spectra.  The lower panels show the fit residuals.}
\label{fig:localspectra}
\end{figure}

\begin{figure}
\FigureFile(\linewidth,\linewidth){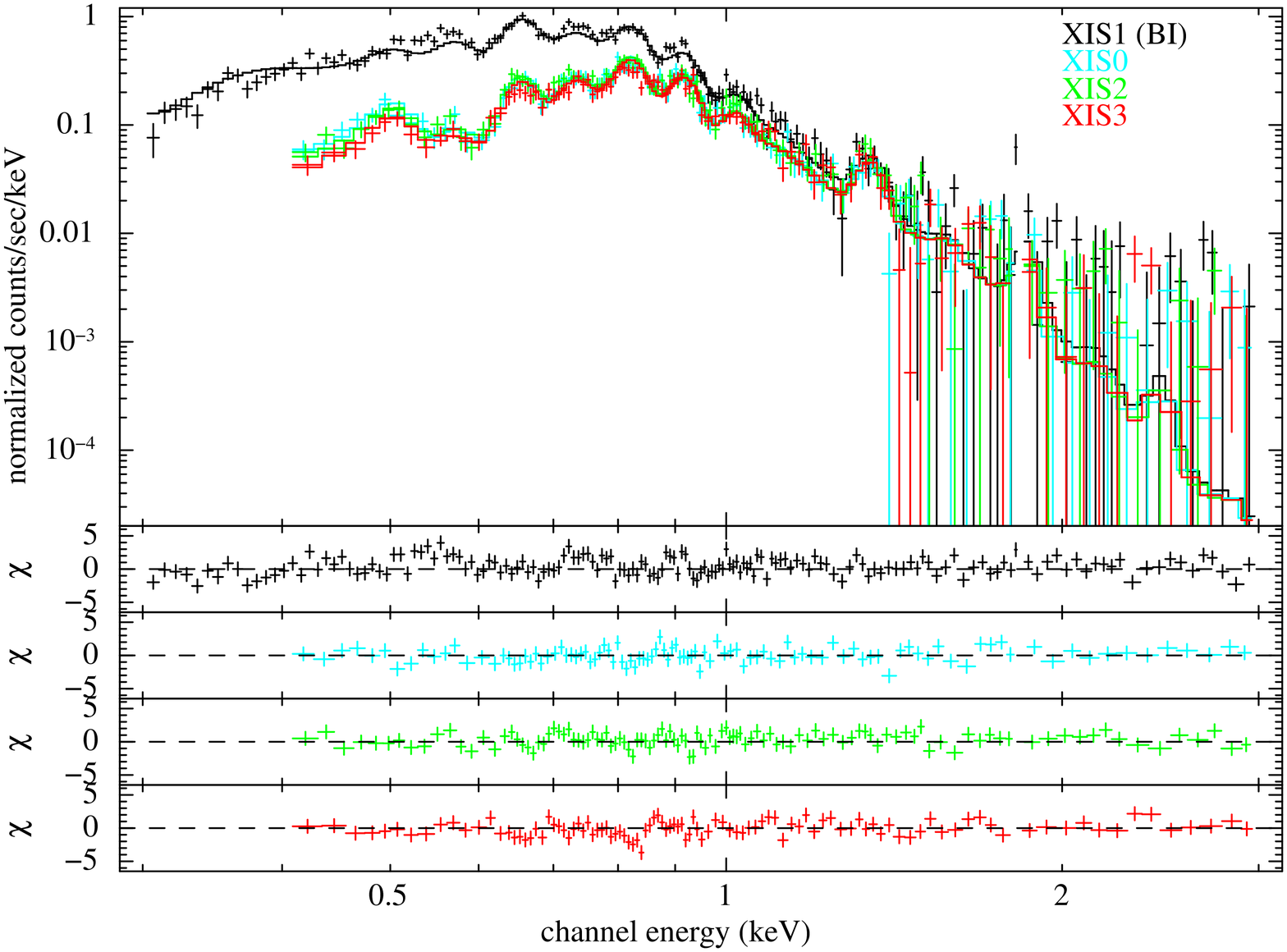}
\caption
{{\it Suzaku\/}/XIS spectra of the NPS overlaid with the best-fit
off-source background model.  The particle background has been subtracted
from the count rate spectra.  Notations are identical to Figure
\ref{fig:localspectra}.}
\label{fig:nepspectra}
\end{figure}

The best-fit parameters obtained with these two background models are
presented in Table \ref{tab:models}, and spectra overlaid with the models
are shown in Figures \ref{fig:localspectra} and \ref{fig:nepspectra}.  The
formal fit is poor for both models, with $\chi_r^2 = 1.21$--$1.24$.  The
residuals in Figures \ref{fig:localspectra} and \ref{fig:nepspectra}
suggest that much of the discrepancy could be from incorrect modeling of
the line spread function in the instrument response, as there are marked
residuals in the wings of strong emission lines.  The XIS1
residuals are larger than those for FI spectra; in particular,
there is an excess of flux compared with the predicted model near 0.54 eV
in the XIS1 spectrum, on the low energy side of the O~\textsc{vii} line.
Also, the Fe~\textsc{xvii} complex near 0.72 keV is brighter in the XIS1
spectrum than the model and FI spectra suggest.  These differences are
possibly due to errors in the early response calibration for the BI sensor.
The general continuum shape and other emission line fluxes appear
to be satisfactorily modeled for both the BI and FI spectra. 

\begin{figure}
\FigureFile(\linewidth,\linewidth){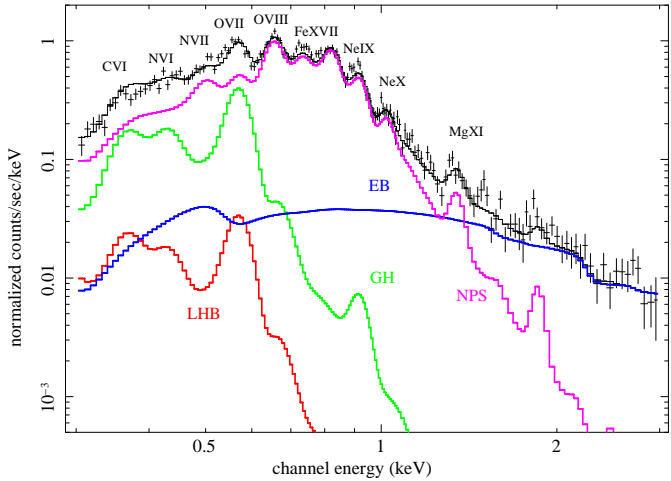}
\caption
{{\it Suzaku\/}/XIS1 spectrum of the NPS, showing the additive emission
components included in the local background model.  Bright emission lines
are marked with the species of the dominant transition responsible for the
emission.  The LHB component is plotted assuming the EM equals the 90\%
upper limit.  The NPS plasma dominates the flux below 1.5 keV, producing
the bright emission lines of N~\textsc{vii} and O~\textsc{viii}, as well as
lines from Fe, Ne and Mg.  The Galactic halo component contributes to the
C~\textsc{vi}, N~\textsc{vi}, and O~\textsc{vii} lines.  The N~\textsc{vii}
line at 0.50 keV can only be produced by the hotter NPS plasma in this
model.}
\label{fig:speccomps}
\end{figure}

In the local background model, different line-of-sight components dominate
the emission at different energies.  This can be seen in Figure
\ref{fig:speccomps}, which shows the BI (XIS1) spectrum along
with individual additive spectral components.  Strong emission lines from
C, N, O, Fe, Ne and Mg are clearly seen, with the majority of these lines
(and the majority of the flux below 1.5 keV) produced by the NPS component.
Notable exceptions are the C~\textsc{vi}, N~\textsc{vi} and O~\textsc{vii}
lines, which are produced largely by the Galactic halo component.  The
N~\textsc{vii} emission line at 0.50 keV is seen toward the NPS for the
first time, and in our model it is produced by the NPS plasma, since the
other thermal components are too cool for this ionization state to occur
under collisional ionization equilibrium (CIE) conditions.

The two background models yield similar best-fit NPS temperature ($kT =
0.26$--$0.29$ keV), remarkable given the very different methods of
accounting for the SXRB.  Both fits suggest low abundances (about 0.2--0.5
solar) for the NPS plasma, although the off-source background model
produces values half as large as the local background model.  (Note we use
the solar abundance tables of \citet{AndersGrevesse1989} to facilitate
comparison to previous studies.) This difference highlights a difficulty of
constraining the abundance of optically thin thermal plasmas at
temperatures of 1--3$\times10^{6}$ K using X-ray data.  At these
temperatures, the spectrum is dominated by bright emission lines from
metals, and the continuum produced primarily by hydrogen and helium
contributes only a few percent of the flux.  Systematic uncertainties in
the cosmic X-ray (AGN) and particle background limit the accuracy with
which the continuum can be determined.  At the low spectral resolution
afforded by CCDs, uncertainties in the wings of the line redistribution
function produce further errors at a flux level comparable to the
continuum.  The resulting poor constraints on the hydrogen number density
yield poor constraints on metal abundance with respect to hydrogen, even
though the metal emission line fluxes may be well-determined.

A more robust tracer of enrichment in low-temperature X-ray gas is the
relative metal abundance.  We have determined abundance ratios with respect
to oxygen for each element in the NPS plasma.  These are listed in Table
\ref{tab:ratios} in terms of the solar abundance ratio from
\citet{AndersGrevesse1989}.  
Since best-fit absolute abundances are expected to be correlated and might
not follow a normal distribution, we estimated abundance ratio confidence
intervals with the Markov chain Monte Carlo (MCMC) simulator implemented in
XSPEC v12.3.1, which employs the Metropolis-Hastings algorithm. We ran 20
parallel chains of 4000 steps each, throwing out the first 2000 steps and
retaining the last 2000 steps for each chain.  Random starting points were
drawn from the multivariate normal approximation to the distribution
produced in the original model fit.  The temperature, individual metal
abundances, and foreground absorption of the NPS plasma were free
parameters for both background models.  The normalization of the GH
component was also allowed as a free parameter for the local background
model.  The Rubin-Gelman convergence parameter $\sqrt{R}$
\citep{GelmanRubin1992} satisfied $\sqrt{R} < 1.2$ for all parameters
varied in the simulation, as required for convergence to an acceptably fair
sample of the model parameter distribution.  The results from all 20 chains
were combined into a single distribution, and central 90\% confidence
intervals were determined for each abundance ratio listed in Table
\ref{tab:ratios}.  
Compared to oxygen, the elements neon, magnesium, and iron are overabundant
by factors of 1.4--2.1 in both the local and off-source models, while
carbon is underabundant by a factor of about 5.  Surprisingly, nitrogen is
overabundant by a factor of $4.0^{+0.4}_{-0.5}$ (local background) and
$3.4^{+0.5}_{-0.3}$ (off-source background) compared to oxygen.  

\begin{table}
\begin{center}
\caption{NPS Abundance Ratios$^a$ \label{tab:ratios}}
\begin{tabular}{ccc}
\hline
\hline
     & local                   & off-source \\
     & background              & background \\
\hline
C/O  & $<$0.17 & $<$0.23 \\
N/O  & 3.98 $^{+0.43}_{-0.45}$ &  3.41 $^{+0.48}_{-0.29}$ \\
Ne/O & 1.53 $^{+0.13}_{-0.11}$ &  1.68 $^{+0.14}_{-0.11}$ \\
Mg/O & 1.39 $^{+0.29}_{-0.24}$ &  2.05 $^{+0.37}_{-0.37}$ \\
Fe/O & 1.50 $^{+0.14}_{-0.12}$ &  2.07 $^{+0.15}_{-0.15}$ \\
\hline
\multicolumn{3}{l}{\parbox{60mm}{\footnotesize
\footnotemark[$a$] Abundance ratios are shown in terms of the solar values
from \citet{AndersGrevesse1989}.
}}
\end{tabular}
\end{center}
\end{table}

\vspace*{1mm}
\subsubsection{Sources of Additional Emission \label{sect:diffuseother}}

The constraints on the NPS nitrogen abundance come primarily from the
N~\textsc{vii} emission line.  The excess of counts near 0.54 keV in the
XIS1 spectrum bears consideration, since this is close to the energy of the
N~\textsc{vii} line.  This excess is also seen in an XIS1 spectrum of the
Carina Nebula \citep{Hamaguchietal2007}, and the authors model it as a
narrow Gaussian line at an energy of $0.547\pm0.01$ keV.  We do the same,
adding a zero-width Gaussian component to the local background model and
freezing all parameters at their best-fit values, allowing only the line
energy and normalization to vary.  Only the XIS1 data are included in the
fit.  The resultant line energy is $0.54\pm0.02$ keV, in agreement with
\citet{Hamaguchietal2007}.  This energy is consistent with fluorescent
scattering of solar X-rays from atmospheric oxygen, however the excess
count rate of 2.0 cts s$^{-1}$ keV$^{-1}$ is much greater than the 
scattered flux of 0.02 cts s$^{-1}$ keV$^{-1}$ estimated in Section
\ref{sect:scatter}.  No other emission lines are expected at this
energy.  It is possible this excess is caused by a feature in the XIS1
response that has not been included in the calibration.  In any case, the
N~\textsc{vii} line is seen in each of the FI spectra, which combined have
an effective area 60\% that of the BI at 0.5 keV.  A fit performed on just
the FI data yields similar results for N/O ratio, and we conclude the 0.54
keV excess has little effect on our results, including the nitrogen
overabundance.

\begin{figure}
\FigureFile(\linewidth,\linewidth){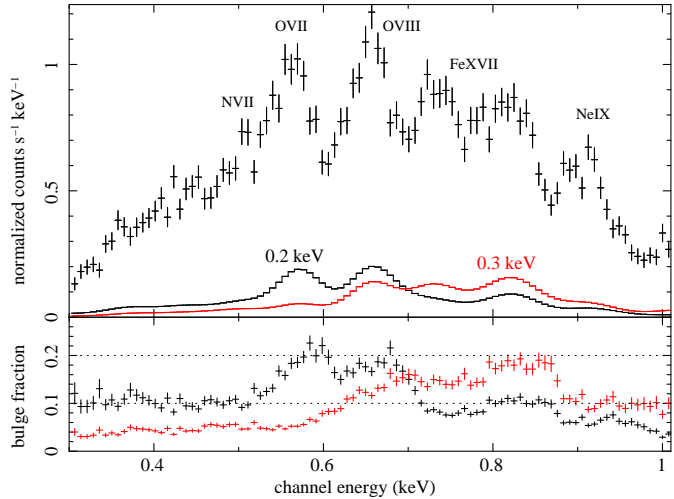}
\caption
{Estimate of the contaminating flux from the Galactic bulge.
The top panel shows the {\it Suzaku\/}/XIS1 spectrum of the NPS (points)
plotted with the modeled Galactic X-ray bulge spectrum (solid lines) along
this line of sight, assuming plasma with temperature 0.2 keV (black) and
0.3 keV (red) from the \protect{\citet{Almyetal2000}} polytrope model.  The
bottom panel shows the fraction of the NPS spectrum attributable to the
X-ray bulge model for each assumed temperature, with dotted lines marking
10\% and 20\%. The bulge emission accounts for up to 20\% of the flux in
the brightest lines and 10\% or less in the remainder of the spectrum.}
\label{fig:bulge}
\end{figure}

Previous studies using broad-band X-ray spectroscopy suggest a higher
temperature of $kT \approx 0.2$ keV for the GH component (e.g.,
\cite{Smithetal2007,Galeazzietal2007}).  Indeed, a careful analysis of the
{\it ROSAT\/} results suggests a two-temperature model for the so-called
``transabsorption emission'', with $kT_{\rm low} \approx 0.1$ keV and
$kT_{\rm hi} \approx 0.25$ keV \citep{KuntzSnowden2000}.  The NPS
temperature derived by \citet{Willingaleetal2003} is similar to that of the
hot GH component, although they point out that the NPS emission is
significantly brighter than the \citet{KuntzSnowden2000} high-latitude
emission within the observed band, and they ignore any contribution from
the GH to the fitted NPS emission.  
A $kT = 0.2$--0.3 keV component to the 
Galactic halo
could produce a
bright N~\textsc{vii} emission line as well.  Such material in CIE would
also require enhanced N/O to explain the observed line ratios.  In the few
observations sensitive to N~\textsc{vii} emission from diffuse Galactic gas,
this is not found
\citep{McCammonetal2002,Fujimotoetal2007,Smithetal2007}.
As a result, the presence of a hotter GH component would dilute the
considerably stronger NPS emission, and it would {\it increase\/} the
derived N/O ratio.

Additional contaminating flux may arise from Galactic bulge plasma, which
has a temperature similar to that of the NPS \citep{Snowdenetal1997}.  Flux
from the bulge would not be removed by subtracting the far-off-source NEP
background, and could mimic emission from the NPS.  The distribution of
this material has been modeled by several authors
\citep{Snowdenetal1997,Almyetal2000}; to estimate the integrated flux along
our sightline, we adopt the polytrope model of \citet{Almyetal2000}.  At a
projected line-of-sight distance of 35\arcdeg\ from the Galactic center, we
expect a {\it ROSAT\/} 3/4 keV count rate of $100\times10^{-6}$ cts s$^{-1}$
arcmin$^{-2}$ from the Galactic bulge plasma (\cite{Almyetal2000}, Figure
6).  Following the emission modeling of \citet{Almyetal2000}, we produced a
number of Raymond-Smith plasma models with this count rate (scaled to the
observed XIS field of view) at a variety of temperatures in the expected
range of 0.2--0.3 keV, folding through the XIS1 response.  The results for
0.2 and 0.3 keV are presented in Figure \ref{fig:bulge}, which shows the
fraction of the observed XIS1 spectrum attributable to the modeled bulge.
For 0.2 keV, the bulge accounts for up to 20\% of the flux in the
O~\textsc{vii} and O~\textsc{viii} lines, and 10\% or less of the flux
elsewhere, including near the N~\textsc{vii} line.  For a 0.3 keV plasma,
the bulge accounts for 20\% of the flux in the O~\textsc{viii} and Fe
lines, with less than 5\% contamination at other energies.  Intermediate
temperatures contribute smaller fractions of the observed line flux.

A recent {\it Suzaku\/}/XIS observation towards a much brighter region of
the X-ray-emitting bulge implies a somewhat lower contaminating flux.  This
region, towards $(l,b) = (1.3\arcdeg,-14.5\arcdeg)$, is about 4 times
brighter in the {\it ROSAT\/} 3/4 keV map compared to the background region
near the NPS.  The XIS spectrum toward this region is 50\% of the flux of
our NPS spectrum in the oxygen emission lines (McCammon et al.~in
preparation).  This implies a bulge contamination of about 10--15\% of the
total observed flux.  The bulge spectrum also shows no N~\textsc{vii}
emission line.  We conclude that while the Galactic bulge might produce a
small fraction of the soft flux seen toward the NPS, it is unable to
explain the enhanced N/O abundance ratio.

A single temperature for the NPS emission might be too simple, and the N/O
ratio could be explained by multiple temperature components.
\citet{Willingaleetal2003} obtain an unusually high EM for the low-$kT$ GH
component compared to other sightlines, as do we, and this is largely due
to the very bright O~\textsc{vii} line.  It is impossible to produce
O~\textsc{vii} along with O~\textsc{viii} and the higher energy lines in
the flux ratios observed with a single-temperature model.  However, it is
reasonable to believe that our line of sight probes multiple temperature
components within the NPS, and that some of the O~\textsc{vii} could be
produced in cooler NPS plasma rather than the GH.  We have investigated the
effect of additional NPS temperature components on our fitting results.
This does not change the quantitative fit, nor does it account for the flux
of the N~\textsc{vii} line without invoking a super-solar N/O ratio in the
NPS plasma.  We conclude that the enhanced nitrogen abundance observed
toward the NPS is real and not a result of superposed, multi-temperature
emission regions.


\section{Discussion}

The NPS plasma conditions presented here are similar to the recent 
{\it XMM-Newton\/} results reported by \citet{Willingaleetal2003} for three
different pointings.  Our temperature is slightly higher at 
$kT = 0.29\pm0.01$ keV compared to 0.25--0.27 keV for {\it XMM-Newton\/}
(90\% confidence error bars).  We obtain a similar intervening absorption
column as \citet{Willingaleetal2003}, supporting the presence of a dense
neutral sheet between the LHB and NPS bubbles.  From the NPS emission
parameters and a suitable model for the emission morphology, we can place
constraints on the physical conditions of the plasma.  We use the schematic
model of \citet{Willingaleetal2003}, which consists of a bubble centered on
$l = 352\arcdeg$, $b = +15\arcdeg$ at a distance of 210 pc with radius 140
pc.  This structure is offset from and smaller than the Loop I bubble
centered on the Sco-Cen association, and it is constructed to correspond to
the bulk of the bright X-ray emission.  Our pointing probes a path length
of 157 pc through the NPS bubble.  For an EM of 0.10 pc cm$^{-6}$, the
electron density along the line of sight is $n_e = 0.028 f^{-1/2}$
cm$^{-3}$, where $f$ is the volume filling factor and assuming uniform
density within regions occupied by the hot gas.  We also assume a fully
ionized plasma with the best-fit abundances from our local background
model, including a solar helium abundance.  Given a best-fit temperature of
0.29 keV, the pressure from electrons and ions is $P/k = 1.8\times 10^{5}
f^{-1/2}$ cm$^{-3}$ K.  These values are consistent with the {\it XMM\/}
results.  The total mass of hot gas in the NPS bubble is $9.3\times 10^{3}
f^{1/2}$ $M_{\odot}$, yielding a total thermal energy of $1.2\times 10^{52}
f^{1/2}$ erg.  This is consistent with the energy output of a single
supernova only if the filling factor of the gas is $f \lesssim 0.01$, much
lower than expected for an evolved SNR.  However, we note that most of
the bubble has a much lower X-ray brightness than this field, therefore the
density and energy derived here should be considered upper limits.  The
results are consistent with a bubble formed by one or more supernovae.

The abundances of O, Ne, Mg and Fe are consistent with previous results
\citep{Willingaleetal2003} and are significantly subsolar, with O/H
$\approx$ 0.3 solar compared to the \citet{AndersGrevesse1989} standard.
Subsolar abundances are observed in other X-ray emission observations of
the ISM (e.g., \cite{Miyataetal2007}) as well as UV and X-ray absorption
measurements
\citep{Meyeretal1997,Meyeretal1998,Takeietal2002,Andreetal2003,YaoWang2006},
and these fall in line with recent revisions in the standard solar
abundance values \citep{Asplundetal2005}.  Our derived O abundance is still
about 50--60\% of the revised solar value, though we have noted previously
the difficulties of obtaining absolute metal abundances from X-ray emission
data and stress that these values are not well-constrained.  The ratios of
Ne/O, Mg/O, and Fe/O are robust tracers of enrichment and are enhanced at
1.4--1.5 times the solar ratio.

The derived abundances of N and C are very different than those of other
metals, and these are new results of this work.  
We observe an enhancement of N toward the NPS, with N/H =
$1.3^{+0.5}_{-0.4}$ solar and N/O = $4.0^{+0.4}_{-0.5}$ solar.  
This is consistent with previous constraints:  early X-ray observations
suggest unusually strong N~\textsc{vi} or N~\textsc{vii} emission for the
assumed NPS thermal model, although the limits are quite large
\citep{Inoueetal1980,Rocchiaetal1984}.  The fields of view of these
observations are also much larger than the {\it Suzaku\/}/XIS.
\citet{Cawley1998} used rocket-based CCD observations and obtained a
best-fit N/O ratio of about 3, but with ``unconstrained'' error bars.  Ours
is the first clear detection of a N~\textsc{vii} emission line toward the
NPS.

We detect a C~\textsc{vi} line in the XIS1 spectrum, but our best-fit local
background model suggests it arises in the GH and LHB, not the NPS.  This
places an upper limit on the NPS abundance of C/H $< 0.06$ solar and C/O $<
0.2$ solar (limits are 90\% confidence).  The robustness of this results
hinges on the nature of the GH component; at 0.1 keV, a solar abundance
plasma must produce very strong C~\textsc{vi} emission to also account for
the bright O~\textsc{vii} line.  If there is a cooler NPS component with
$kT \approx 0.2$ keV, as discussed earlier, this could account for much of
the O~\textsc{vii} emission without filling in the C~\textsc{vi} line
completely, thus allowing some contribution from the hotter 0.3 keV NPS
gas.  The similar result for the local background and off-source
background models suggests that the GH accounts for the full C~\textsc{vi}
and that the C/O deficit is real.

We can shed light on the N/O enhancement and possible C/O depletion by
considering the mechanisms that produce these metals.  In the
generally-accepted view of stellar metal enrichment, carbon and oxygen are
produced primarily by helium shell burning in massive stars and released to
the ISM through supernovae.  Nitrogen, on the other hand, is produced
largely by intermediate-mass (4--8 $M_{\odot}$) stars and injected into the
ISM via AGB winds \citep{Henryetal2000}.  Carbon and oxygen are effectively
converted to nitrogen during the CN branch of the CNO cycle, in which
destruction of $^{14}$N is the limiting process
($^{14}$N(p,$\gamma$)$^{15}$O; e.g., \cite{Clayton1983}).  In AGB stars,
this process occurs in the hydrogen-burning shell, with convection acting
to cycle unprocessed envelope material through the CNO burning region and
dredge-up returning this nitrogen-enriched material to the surface where it
is released to the ISM in a strong stellar wind
\citep{Scaloetal1975,IbenRenzini1983}.  Evidence for large nitrogen
enhancements and carbon deficiencies have been seen in AGB star envelopes
\citep{McSaveneyetal2007}.  High mass evolved stars such as Wolf-Rayets and
Luminous Blue Variables can also enrich their surroundings with nitrogen,
although the effects of this appear to be small \citep{Henryetal2000}.

The large nitrogen overabundance observed in the NPS suggests enrichment by
AGB activity of the X-ray emitting material, yet this is difficult to
reconcile with the conventional view of the NPS origin.  In the superbubble
view, the combined effects of stellar winds and occasional supernovae from
the Scorpius-Centaurus OB association have swept out a shell of hot
material encompassed by the Loop I radio emission
\citep{deGeus1992,EggerAschenbach1995}.  The bright X-ray emission is
produced by the most recent supernova shock wave (about $10^{5}$ years old)
heating the inner shell of the Loop I bubble, with an off-center progenitor
star explaining the asymmetry \citep{EggerAschenbach1995}.  Photometric
analysis of the Sco-Cen members \citep{deGeusetal1989} and the measured
expansion rate of the H~\textsc{i} shell
\citep{Sofueetal1974,Heilesetal1980} constrain the age of the superbubble
to be $10^{6}$--$10^{7}$ yr, less than the $10^{8}$ yr required for the
first appearance of intermediate-mass AGB stars \citep{Henryetal2000}.  It
is therefore unlikely that the Sco-Cen association is responsible for the
nitrogen enhancement in the reheated NPS plasma.

It is possible that the metal distribution of the NPS material was altered
prior to formation of the Sco-Cen association and the Loop I superbubble.
Observations of O~\textsc{i} and N~\textsc{i} absorption against stars using
{\it HST\/} and {\it FUSE\/} suggest a systematic trend of lower nitrogen
abundance and lower N/O ratio with increasing total hydrogen column density
\citep{Meyeretal1997,Andreetal2003,Knauthetal2003}.  Most notably, a recent
reanalysis of these data shows a sharp drop in the ISM N/O ratio at a
distance of 500 pc, with N/O $\approx$ 1.7 solar within this distance and
near solar beyond it \citep{Knauthetal2006}.  The NPS bubble as modeled
lies fully within 500 pc radius, and while there is a discrepancy in the
values of N/O, the trend of enhanced N/O ratio we infer is similar to that
seen in the UV absorption observations.

Two possible mechanisms for this metallicity inhomogeneity are suggested by
\citet{Knauthetal2006}.  First, incomplete mixing of AGB wind and
supernova products can alter the local abundance distribution for a short
time.  The observed N/O overabundance could result from enhanced AGB
activity in the local Galaxy during the recent past.  Second, localized
infall of low-metallicity (perhaps primordial) gas could alter the
abundance pattern, initially diluting all metals, but eventually producing
a super-solar N/O ratio through enhanced intermediate-mass star production
\citep{KoppenHensler2005}.  Further investigation is necessary to
confirm the abundance inhomogeneity in different temperature phases and 
to identify its origin.

\section{Summary}

We have presented {\it Suzaku\/}/XIS observations of the North Polar Spur,
observing the N~\textsc{vii} emission line at 0.50 keV for the first time
toward this sightline.  Using two methods to account for the X-ray
background, and assuming CIE, we determine the NPS plasma is best fit by a
thermal APEC model with $kT = 0.29\pm0.01$ keV and C, O, Ne, Mg, and Fe
abundances of $< 0.5$ solar.  Nitrogen appears enhanced, with a best-fit
N/O abundance ratio of $4.0^{+0.4}_{-0.5}$ times the solar ratio.  The
temperature and total thermal energy of the gas suggest heating by multiple
supernovae, consistent with previous models for the NPS and Loop I emission
\citep{EggerAschenbach1995}.  The enhanced nitrogen abundance is best
explained by enrichment from stellar material that has been processed by
the CNO cycle, likely the result of enhanced AGB activity.  Due to the
time required to develop AGB stars, we conclude that this N/O enhancement
cannot be caused by the Sco-Cen OB association, but may result from a
previous enrichment episode in the solar neighborhood.

\vspace{\baselineskip}
We owe a debt of gratitude to the entire {\it Suzaku\/} team for their
support of the spacecraft and science operations.  We thank Randall Smith,
Norbert Nemes, and Eric Feigelson for valuable input, and thank the referee
for useful comments that improved the manuscript.  We also thank the ACE
SWEPAM instrument team and the ACE Science Center for providing the ACE
data.  EDM acknowledges support from a Short-Term Postdoctoral Fellowship
granted by the Japan Society for the Promotion of Science (JSPS).  This
work is financially supported by NASA grant NNG05GM92G and by a
Grant-in-Aid for Scientific Research by the Ministry of Education, Culture,
Sports, Science and Technology of Japan (16002004).



\begin{thebibliography}{63}
\expandafter\ifx\csname natexlab\endcsname\relax\def\natexlab#1{#1}\fi

\bibitem[{{Almy} {et~al.}(2000){Almy}, {McCammon}, {Digel}, {Bronfman}, \&
  {May}}]{Almyetal2000}
{Almy}, R.~C., {McCammon}, D., {Digel}, S.~W., {Bronfman}, L., \& {May}, J.
  2000, \apj, 545, 290

\bibitem[{{Anders} \& {Grevesse}(1989)}]{AndersGrevesse1989}
{Anders}, E. \& {Grevesse}, N. 1989, \gca, 53, 197

\bibitem[{{Andr{\'e}} {et~al.}(2003){Andr{\'e}}, {Oliveira}, {Howk}, {Ferlet},
  {D{\'e}sert}, {H{\'e}brard}, {Lacour}, {Lecavelier des {\'E}tangs},
  {Vidal-Madjar}, \& {Moos}}]{Andreetal2003}
{Andr{\'e}}, M.~K. et al.  2003, \apj, 591, 1000

\bibitem[{{Arnaud}(1996)}]{Arnaud1996}
{Arnaud}, K.~A. 1996, in ASP Conf.~Series, Vol.~101, Astronomical Data
  Analysis Software and Systems V, ed. G.~H. {Jacoby} \& J.~{Barnes}, 17

\bibitem[{{Asplund} {et~al.}(2005){Asplund}, {Grevesse}, \&
  {Sauval}}]{Asplundetal2005}
{Asplund}, M., {Grevesse}, N., \& {Sauval}, A.~J. 2005, in ASP
  Conf.~Series, Vol.~336, Cosmic Abundances as Records of
  Stellar Evolution and Nucleosynthesis, ed. T.~G. {Barnes}, III \& F.~N.
  {Bash}, 25

\bibitem[{{Bauer} {et~al.}(2004){Bauer}, {Alexander}, {Brandt}, {Schneider},
  {Treister}, {Hornschemeier}, \& {Garmire}}]{Baueretal2004}
{Bauer}, F.~E., {Alexander}, D.~M., {Brandt}, W.~N., {Schneider}, D.~P.,
  {Treister}, E., {Hornschemeier}, A.~E., \& {Garmire}, G.~P. 2004, \aj, 128,
  2048

\bibitem[{{Berkhuijsen}(1971)}]{Berkhuijsen1971}
{Berkhuijsen}, E.~M. 1971, \aap, 14, 359

\bibitem[{{Berkhuijsen} {et~al.}(1971){Berkhuijsen}, {Haslam}, \&
  {Salter}}]{Berkhuijsenetal1971}
{Berkhuijsen}, E.~M., {Haslam}, C.~G.~T., \& {Salter}, C.~J. 1971, \aap, 14,
  252

\bibitem[{{Bland-Hawthorn} \& {Cohen}(2003)}]{Bland-HawthornCohen2003}
{Bland-Hawthorn}, J. \& {Cohen}, M. 2003, \apj, 582, 246

\bibitem[{{Bunner} {et~al.}(1972){Bunner}, {Coleman}, {Kraushaar}, \&
  {McCammon}}]{Bunneretal1972}
{Bunner}, A.~N., {Coleman}, P.~L., {Kraushaar}, W.~L., \& {McCammon}, D. 1972,
  \apj, 172, L67

\bibitem[{{Cannon} \& {Pickering}(1918)}]{HDCat}
{Cannon}, A. \& {Pickering}, E. 1918, Annals of Harvard College Observatory, 91

\bibitem[{{Cawley}(1998)}]{Cawley1998}
{Cawley}, L.~J. 1998, PhD thesis, The Pennsylvania State University

\bibitem[{{Centurion} \& {Vladilo}(1991)}]{CenturionVladilo1991}
{Centurion}, M. \& {Vladilo}, G. 1991, \apj, 372, 494

\bibitem[{{Clayton}(1983)}]{Clayton1983}
{Clayton}, D.~D. 1983, {Principles of stellar evolution and nucleosynthesis}
  (Chicago: University of Chicago Press)

\bibitem[{{Cravens}(2000)}]{Cravens2000}
{Cravens}, T.~E. 2000, \apj, 532, L153

\bibitem[{{Cruddace} {et~al.}(1976){Cruddace}, {Friedman}, {Fritz}, \&
  {Shulman}}]{Cruddaceetal1976}
{Cruddace}, R.~G., {Friedman}, H., {Fritz}, G., \& {Shulman}, S. 1976, \apj,
  207, 888

\bibitem[{{Davelaar} {et~al.}(1980){Davelaar}, {Bleeker}, \&
  {Deerenberg}}]{Davelaaretal1980}
{Davelaar}, J., {Bleeker}, J.~A.~M., \& {Deerenberg}, A.~J.~M. 1980, \aap, 92,
  231

\bibitem[{{de Geus}(1992)}]{deGeus1992}
{de Geus}, E.~J. 1992, \aap, 262, 258

\bibitem[{{de Geus} {et~al.}(1989){de Geus}, {de Zeeuw}, \&
  {Lub}}]{deGeusetal1989}
{de Geus}, E.~J., {de Zeeuw}, P.~T., \& {Lub}, J. 1989, \aap, 216, 44

\bibitem[{{De Luca} \& {Molendi}(2004)}]{DeLucaMolendi2004}
{De Luca}, A. \& {Molendi}, S. 2004, \aap, 419, 837

\bibitem[{{Dickey} \& {Lockman}(1990)}]{DickeyLockman1990}
{Dickey}, J.~M. \& {Lockman}, F.~J. 1990, \araa, 28, 215

\bibitem[{{Drake} {et~al.}(1989){Drake}, {Simon}, \& {Linsky}}]{Drakeetal1989}
{Drake}, S.~A., {Simon}, T., \& {Linsky}, J.~L. 1989, \apjs, 71, 905

\bibitem[{{Egger}(1995)}]{Egger1995}
{Egger}, R.~J. 1995, in ASP Conf.~Series,
  Vol.~80, The Physics of the Interstellar Medium and Intergalactic Medium, ed.
  A.~{Ferrara}, C.~F. {McKee}, C.~{Heiles}, \& P.~R. {Shapiro}, 45

\bibitem[{{Egger} \& {Aschenbach}(1995)}]{EggerAschenbach1995}
{Egger}, R.~J. \& {Aschenbach}, B. 1995, \aap, 294, L25

\bibitem[{{Fujimoto} {et~al.}(2007){Fujimoto}, {Mitsuda}, {Mccammon}, {Takei},
  {Bauer}, {Ishisaki}, {Porter}, {Yamaguchi}, {Hayashida}, \&
  {Yamasaki}}]{Fujimotoetal2007}
{Fujimoto}, R. et al.  2007, \pasj, 59, 133

\bibitem[{{Galeazzi} {et~al.}(2007){Galeazzi}, {Gupta}, {Covey}, \&
  {Ursino}}]{Galeazzietal2007}
{Galeazzi}, M., {Gupta}, A., {Covey}, K., \& {Ursino}, E. 2007, \apj, 658, 1081

\bibitem[{{Gelman} \& {Rubin}(1992)}]{GelmanRubin1992}
{Gelman}, A. \& {Rubin}, D.~B. 1992, Stat.~Sci., 7, 457

\bibitem[{{Hamaguchi} {et~al.}(2007){Hamaguchi}, {Petre}, {Matsumoto},
  {Tsujimoto}, {Holt}, {Ezoe}, {Ozawa}, {Tsuboi}, {Soong}, {Kitamoto},
  {Sekiguchi}, \& {Kokubun}}]{Hamaguchietal2007}
{Hamaguchi}, K. et al.  2007, \pasj, 59, 151

\bibitem[{{Heiles} {et~al.}(1980){Heiles}, {Chu}, {Troland}, {Reynolds}, \&
  {Yegingil}}]{Heilesetal1980}
{Heiles}, C., {Chu}, Y.-H., {Troland}, T.~H., {Reynolds}, R.~J., \& {Yegingil},
  I. 1980, \apj, 242, 533

\bibitem[{{Henry} {et~al.}(2000){Henry}, {Edmunds}, \&
  {K{\"o}ppen}}]{Henryetal2000}
{Henry}, R.~B.~C., {Edmunds}, M.~G., \& {K{\"o}ppen}, J. 2000, \apj, 541, 660

\bibitem[{{Houk} \& {Swift}(2000)}]{Houk}
{Houk}, N. \& {Swift}, C. 2000, {Michigan Catalogue of HD stars, Vol.~5}

\bibitem[{{Iben} \& {Renzini}(1983)}]{IbenRenzini1983}
{Iben}, Jr., I. \& {Renzini}, A. 1983, \araa, 21, 271

\bibitem[{{Inoue} {et~al.}(1980){Inoue}, {Koyama}, {Matsuoka}, {Ohashi},
  {Tanaka}, \& {Tsunemi}}]{Inoueetal1980}
{Inoue}, H., {Koyama}, K., {Matsuoka}, M., {Ohashi}, T., {Tanaka}, Y., \&
  {Tsunemi}, H. 1980, \apj, 238, 886

\bibitem[{{Ishisaki} {et~al.}(2007){Ishisaki}, {Maeda}, {Fujimoto}, {Ozaki},
  {Ebisawa}, {Takahashi}, {Ueda}, {Ogasaka}, {Ptak}, {Mukai}, {Hamaguchi},
  {Hirayama}, {Kotani}, {Kubo}, {Shibata}, {Ebara}, {Furuzawa}, {Iizuka},
  {Inoue}, {Mori}, {Okada}, {Yokoyama}, {Matsumoto}, {Nakajima}, {Yamaguchi},
  {Anabuki}, {Tawa}, {Nagai}, {Katsuda}, {Hayashida}, {Bamba}, {Miller},
  {Sato}, \& {Yamasaki}}]{Ishisakietal2007}
{Ishisaki}, Y. et al.  2007, \pasj, 59, 113

\bibitem[{{Iwan}(1980)}]{Iwan1980}
{Iwan}, D. 1980, \apj, 239, 316

\bibitem[{{Knauth} {et~al.}(2003){Knauth}, {Andersson}, {McCandliss}, \&
  {Moos}}]{Knauthetal2003}
{Knauth}, D.~C., {Andersson}, B.-G., {McCandliss}, S.~R., \& {Moos}, H.~W.
  2003, \apj, 596, L51

\bibitem[{{Knauth} {et~al.}(2006){Knauth}, {Meyer}, \&
  {Lauroesch}}]{Knauthetal2006}
{Knauth}, D.~C., {Meyer}, D.~M., \& {Lauroesch}, J.~T. 2006, \apj, 647, L115

\bibitem[{{K{\"o}ppen} \& {Hensler}(2005)}]{KoppenHensler2005}
{K{\"o}ppen}, J. \& {Hensler}, G. 2005, \aap, 434, 531

\bibitem[{{Koyama} {et~al.}(2007){Koyama}, {Tsunemi}, {Dotani}, {Bautz},
  {Hayashida}, {Tsuru}, {Matsumoto}, {Ogawara}, {Ricker}, {Doty}, {Kissel},
  {Foster}, {Nakajima}, {Yamaguchi}, {Mori}, {Sakano}, {Hamaguchi},
  {Nishiuchi}, {Miyata}, {Torii}, {Namiki}, {Katsuda}, {Matsuura}, {Miyauchi},
  {Anabuki}, {Tawa}, {Ozaki}, {Murakami}, {Maeda}, {Ichikawa}, {Prigozhin},
  {Boughan}, {Lamarr}, {Miller}, {Burke}, {Gregory}, {Pillsbury}, {Bamba},
  {Hiraga}, {Senda}, {Katayama}, {Kitamoto}, {Tsujimoto}, {Kohmura}, {Tsuboi},
  \& {Awaki}}]{Koyamaetal2007}
{Koyama}, K. et al.  2007, \pasj, 59, 23

\bibitem[{{Kuntz} \& {Snowden}(2000)}]{KuntzSnowden2000}
{Kuntz}, K.~D. \& {Snowden}, S.~L. 2000, \apj, 543, 195

\bibitem[{{Mathewson} \& {Ford}(1970)}]{MathewsonFord1970}
{Mathewson}, D.~S. \& {Ford}, V.~L. 1970, \memras, 74, 139

\bibitem[{{McCammon} {et~al.}(2002){McCammon}, {Almy}, {Apodaca}, {Bergmann
  Tiest}, {Cui}, {Deiker}, {Galeazzi}, {Juda}, {Lesser}, {Mihara},
  {Morgenthaler}, {Sanders}, {Zhang}, {Figueroa-Feliciano}, {Kelley},
  {Moseley}, {Mushotzky}, {Porter}, {Stahle}, \&
  {Szymkowiak}}]{McCammonetal2002}
{McCammon}, D. et al.  2002, \apj, 576, 188

\bibitem[{{McSaveney} {et~al.}(2007){McSaveney}, {Wood}, {Scholz}, {Lattanzio},
  \& {Hinkle}}]{McSaveneyetal2007}
{McSaveney}, J.~A., {Wood}, P.~R., {Scholz}, M., {Lattanzio}, J.~C., \&
  {Hinkle}, K.~H. 2007, ArXiv e-prints, 704

\bibitem[{{Meyer} {et~al.}(1997){Meyer}, {Cardelli}, \&
  {Sofia}}]{Meyeretal1997}
{Meyer}, D.~M., {Cardelli}, J.~A., \& {Sofia}, U.~J. 1997, \apj, 490, L103

\bibitem[{{Meyer} {et~al.}(1998){Meyer}, {Jura}, \& {Cardelli}}]{Meyeretal1998}
{Meyer}, D.~M., {Jura}, M., \& {Cardelli}, J.~A. 1998, \apj, 493, 222

\bibitem[{{Miyata} {et~al.}(2007){Miyata}, {Katsuda}, {Tsunemi}, {Hughes},
  {Kokubun}, \& {Porter}}]{Miyataetal2007}
{Miyata}, E., {Katsuda}, S., {Tsunemi}, H., {Hughes}, J.~P., {Kokubun}, M., \&
  {Porter}, F.~S. 2007, \pasj, 59, 163

\bibitem[{{Picone} {et~al.}(2002){Picone}, {Hedin}, {Drob}, \&
  {Aikin}}]{Piconeetal2002}
{Picone}, J.~M., {Hedin}, A.~E., {Drob}, D.~P., \& {Aikin}, A.~C. 2002, Journal
  of Geophysical Research (Space Physics), 107, 15

\bibitem[{{Raines} {et~al.}(2005){Raines}, {Lepri}, {Zurbuchen}, \& {et
  al.}}]{Rainesetal2005}
{Raines}, J., {Lepri}, S.~T., {Zurbuchen}, T.~H., \& {et al.} 2005, in ESA
  SP-592: Solar Wind 11/SOHO 16, Connecting Sun and Heliosphere, Vol.~16

\bibitem[{{Roberts} \& {Warwick}(2001)}]{RobertsWarwick2001}
{Roberts}, T.~P. \& {Warwick}, R.~S. 2001, in ASP Conf.~Series,
  Vol.~234, X-ray Astronomy 2000, ed. R.~{Giacconi},
  S.~{Serio}, \& L.~{Stella}, 569

\bibitem[{{Rocchia} {et~al.}(1984){Rocchia}, {Arnaud}, {Blondel}, {Cheron},
  {Christy}, {Rothenflug}, {Schnopper}, \& {Delvaille}}]{Rocchiaetal1984}
{Rocchia}, R. et al.  1984, \aap, 130, 53

\bibitem[{{Scalo} {et~al.}(1975){Scalo}, {Despain}, \&
  {Ulrich}}]{Scaloetal1975}
{Scalo}, J.~M., {Despain}, K.~H., \& {Ulrich}, R.~K. 1975, \apj, 196, 805

\bibitem[{{Schnopper} {et~al.}(1982){Schnopper}, {Delvaille}, {Rocchia},
  {Blondel}, {Cheron}, {Christy}, {Ducros}, {Koch}, \&
  {Rothenflug}}]{Schnopperetal1982}
{Schnopper}, H.~W. et al.  \apj, 253, 131

\bibitem[{{Smith} {et~al.}(2007){Smith}, {Bautz}, {Edgar}, {Fujimoto},
  {Hamaguchi}, {Hughes}, {Ishida}, {Kelley}, {Kilbourne}, {Kuntz}, {McCammon},
  {Miller}, {Mitsuda}, {Mukai}, {Plucinsky}, {Porter}, {Snowden}, {Takei},
  {Terada}, {Tsuboi}, \& {Yamasaki}}]{Smithetal2007}
{Smith}, R.~K. et al.  2007, \pasj, 59, 141

\bibitem[{{Smith} {et~al.}(2001){Smith}, {Brickhouse}, {Liedahl}, \&
  {Raymond}}]{Smithetal2001}
{Smith}, R.~K., {Brickhouse}, N.~S., {Liedahl}, D.~A., \& {Raymond}, J.~C.
  2001, \apj, 556, L91

\bibitem[{{Snowden}(1998)}]{Snowden1998}
{Snowden}, S.~L. 1998, \apjs, 117, 233

\bibitem[{{Snowden} {et~al.}(1990){Snowden}, {Cox}, {McCammon}, \&
  {Sanders}}]{Snowdenetal1990}
{Snowden}, S.~L., {Cox}, D.~P., {McCammon}, D., \& {Sanders}, W.~T. 1990, \apj,
  354, 211

\bibitem[{{Snowden} {et~al.}(1997){Snowden}, {Egger}, {Freyberg}, {McCammon},
  {Plucinsky}, {Sanders}, {Schmitt}, {Truemper}, \& {Voges}}]{Snowdenetal1997}
{Snowden}, S.~L. et al.  1997, \apj, 485, 125

\bibitem[{{Snowden} \& {Freyberg}(1993)}]{SnowdenFreyberg1993}
{Snowden}, S.~L. \& {Freyberg}, M.~J. 1993, \apj, 404, 403

\bibitem[{{Sofue}(1977)}]{Sofue1977}
{Sofue}, Y. 1977, \aap, 60, 327

\bibitem[{{Sofue}(1984)}]{Sofue1984}
{Sofue}, Y. 1984, \pasj, 36, 539

\bibitem[{{Sofue}(1994)}]{Sofue1994}
{Sofue}, Y. 1994, \apj, 431, L91

\bibitem[{{Sofue}(2000)}]{Sofue2000}
{Sofue}, Y. 2000, \apj, 540, 224

\bibitem[{{Sofue}(2003)}]{Sofue2003}
{Sofue}, Y. 2003, \pasj, 55, 445

\bibitem[{{Sofue} {et~al.}(1974){Sofue}, {Hamajima}, \&
  {Fujimoto}}]{Sofueetal1974}
{Sofue}, Y., {Hamajima}, K., \& {Fujimoto}, M. 1974, \pasj, 26, 399

\bibitem[{{Stone} {et~al.}(1998){Stone}, {Frandsen}, {Mewaldt}, {Christian},
  {Margolies}, {Ormes}, \& {Snow}}]{ACE}
{Stone}, E.~C., {Frandsen}, A.~M., {Mewaldt}, R.~A., {Christian}, E.~R.,
  {Margolies}, D., {Ormes}, J.~F., \& {Snow}, F. 1998, Space Science Reviews,
  86, 1

\bibitem[{{Takei} {et~al.}(2002){Takei}, {Fujimoto}, {Mitsuda}, \&
  {Onaka}}]{Takeietal2002}
{Takei}, Y., {Fujimoto}, R., {Mitsuda}, K., \& {Onaka}, T. 2002, \apj, 581, 307

\bibitem[{{Tawa} {et~al.}(2007)}]{Tawaetal2007}
{Tawa}, N. et al. 2007, \pasj, submitted

\bibitem[{{Verkhodanov} {et~al.}(2002){Verkhodanov}, {Kopylov}, {Pariiskii},
  {Soboleva}, {Temirova}, \& {Zhelenkova}}]{Verkhodanovetal2002}
{Verkhodanov}, O.~V., {Kopylov}, A.~I., {Pariiskii}, Y.~N., {Soboleva}, N.~S.,
  {Temirova}, A.~V., \& {Zhelenkova}, O.~P. 2002, Astronomy Reports, 46, 531

\bibitem[{{Voges} {et~al.}(1999){Voges}, {Aschenbach}, {Boller},
  {Br{\"a}uninger}, {Briel}, {Burkert}, {Dennerl}, {Englhauser}, {Gruber},
  {Haberl}, {Hartner}, {Hasinger}, {K{\"u}rster}, {Pfeffermann}, {Pietsch},
  {Predehl}, {Rosso}, {Schmitt}, {Tr{\"u}mper}, \&
  {Zimmermann}}]{Vogesetal1999}
{Voges}, W. et al.  1999, \aap, 349, 389

\bibitem[{{Willingale} {et~al.}(2003){Willingale}, {Hands}, {Warwick},
  {Snowden}, \& {Burrows}}]{Willingaleetal2003}
{Willingale}, R., {Hands}, A.~D.~P., {Warwick}, R.~S., {Snowden}, S.~L., \&
  {Burrows}, D.~N. 2003, \mnras, 343, 995

\bibitem[{{Yao} \& {Wang}(2006)}]{YaoWang2006}
{Yao}, Y. \& {Wang}, Q.~D. 2006, \apj, 641, 930

\end{thebibliography}
\end{document}